\documentclass{ptephy_v1}

\preprintnumber{2107-10522} 

\usepackage{amsmath} 
\usepackage{hyperref}
\usepackage{graphics}
\usepackage{bm}

\begin{document}

\title{Rheology of a dilute binary mixture of inertial suspension under simple shear flow}

\author{Satoshi Takada}
\affil{Institute of Engineering and Department of Mechanical Systems Engineering, 
	Tokyo University of Agriculture and Technology,
	2--24--16, Naka-cho, Koganei, Tokyo 184--8588, Japan 
	\email{takada@go.tuat.ac.jp}}

\author{Hisao Hayakawa}
\affil{Yukawa Institute for Theoretical Physics, Kyoto University,
	Kitashirakawa Oiwakecho, Sakyo-ku, Kyoto 606--8502, Japan
	\email{hisao@yukawa.kyoto-u.ac.jp}}

\author{Vicente Garz\'o}
\affil{Departamento de F\'{\i}sica and 
	Instituto de Computaci\'on Cient\'{\i}fica Avanzada (ICCAEX),
	Universidad de Extremadura, E--06006 Badajoz, Spain
	\email{vicenteg@unex.es}}

\begin{abstract}
The rheology of a dilute binary mixture of inertial suspension under simple shear flow is analyzed in the context of the Boltzmann kinetic equation. 
The effect of the surrounding viscous gas on the solid particles is accounted for by means of a deterministic viscous drag force plus a stochastic Langevin-like term defined in terms of the environmental temperature $T_\text{env}$. 
Grad's moment method is employed to determine the temperature ratio and the pressure tensor in terms of the coefficients of restitution, concentration, the masses and diameters of the components of the mixture, and the environmental temperature. 
Analytical results are compared against event-driven Langevin simulations for mixtures of hard spheres with the same mass density $m_1/m_2=(\sigma^{(1)}/\sigma^{(2)})^3$, $m_i$ and $\sigma^{(1)}$ being the mass and diameter, respectively, of the species $i$. 
It is confirmed that the theoretical predictions agree with simulations of various size ratios $\sigma^{(1)}/\sigma^{(2)}$ and for elastic and inelastic collisions in the wide range of parameters' space. 
It is remarkable that the temperature ratio $T_1/T_2$ and the viscosity ratio $\eta_1/\eta_2$ ($\eta_i$ being the partial contribution of the species $i$ to the total shear viscosity $\eta=\eta_1+\eta_2$) discontinuously change at a certain shear rate as the size ratio increases; this feature (which is expected to occur in the thermodynamic limit) cannot be completely captured by simulations due to small system size. 
In addition, a Bhatnagar--Gross--Krook (BGK)-type kinetic model adapted to mixtures of inelastic hard spheres is exactly solved when $T_\text{env}$ is much smaller than the kinetic temperature $T$. 
A comparison between the velocity distribution functions obtained from Grad's method, BGK model, and simulations is carried out.
\end{abstract}

\subjectindex{A57, J01, J44}

\maketitle

\section{Introduction}
\label{sec1}

Rheology is the subject that studies the flow properties of materials. Although the viscosity of the Newtonian fluid is independent of the shear rate, there are many domestic substances (liquids containing microstructures such as suspensions and polymers) where the viscosity depends on the shear rate (non-Newtonian fluids). Within the class of non-Newtonian fluids, some of them exhibit shear thinning (namely, when the viscosity \emph{decreases} with the shear rate) while others display shear thickening (namely, when the viscosity \emph{increases} with the shear rate).
The shear thickening is also categorized into two classes as the continuous shear thickening (CST) and discontinuous shear thickening (DST).
The viscosity increases continuously in CST, while it abruptly changes discontinuously from a small value to a large value at a critical shear rate in DST.
DST has attracted much attention among physicists in the last few years~\cite{Barnes89,Mewis11,Otsuki11,Seto13,Brown14,Lootens05,Cwalina14} as a typical nonequilibrium discontinuous phase transition between a {liquid-like phase and a solid-like phase. In addition, the understanding of the origin of DST is also important for potential industrial applications such as} protective vests and traction controls.

Although most of the previous studies on shear thickening have been oriented to dense suspensions, there are some other studies that analyze a DST-like process for the kinetic temperature of inertial suspensions. {This type of inertial suspensions} can be regarded as an idealistic model of aerosols~\cite{Koch01}.
The DST-like process (or the ignited-quenched transition) of dilute inertial suspensions takes place as a result of a saddle-node bifurcation. {On the other hand}, the DST-like process for dilute suspensions becomes CST-like as the density of suspensions increases~\cite{Tsao95,Sangani96,Chamorro15,BGK2016,DST16,Saha17,Hayakawa17,Takada20}.

To gain some insight into the understanding of the generic features of rheological phase transitions, we use kinetic theory tools in this paper. 
This allows us to offer a quantitative theoretical analysis for the DST-like and CST-like processes in inertial suspensions.
However, it should be noted that some previous kinetic theories for inertial suspensions have ignored thermal fluctuations in the dynamics of grains~\cite{Tsao95,Sangani96,Chamorro15,Saha17}. A refined suspension model including a Langevin-like term has been more recently considered in Refs.\ \cite{BGK2016,DST16,Hayakawa17,Takada20,Sugimoto20}.
The quantitative validity of these studies has already been verified by the event-driven Langevin simulation for hard spheres (EDLSHS)~\cite{Scala12,Hayakawa17,Takada20}.

Most of the previous theoretical studies on the rheology of inertial suspensions have focused on monodisperse systems, namely, suspensions containing only identical spherical particles.
In reality, suspended particles are not identical since the size of the particles is distributed and the shape and mechanical properties of the particles are also different. To quantify the impact of polydispersity on the rheological properties of inertial suspensions under simple or uniform shear flows (USF), we consider a binary mixture in this paper, namely, a suspension which contains two kinds of spherical particles having different sizes.
We note that bidisperse systems are also studied in colloidal and blood suspensions~\cite{Krishnan96, Gamonpilas16, Pednekar18, Liu19}.

A challenging and interesting problem in sheared granular binary mixtures is that of the diffusion. It is well established that in the absence of shear the mass flux is proportional to the density, pressure, and temperature gradients where the corresponding transport coefficients are scalar quantities~\cite{Garzo}. However, when the mixture is strongly sheared, due to the anisotropy induced by the shear flow tensorial quantities are required to characterize the mass transport instead of the conventional scalar diffusion coefficients. There have been some previous attempts for describing the self-diffusion problem in sheared granular mixtures~\cite{Zik91,Menon97}.
As expected, all previous studies indicate that the diffusion process in USF is highly anisotropic and the components of the diffusion can be observed in the directions parallel and perpendicular to the velocity gradient.
To characterize such anisotropy of the diffusion tensor, there have been several theoretical studies based on kinetic theory~\cite{Garzo02,Garzo07d}, simulation works of rapid granular shear flows~\cite{Campbell97,ALJR21}, and experimental studies of dense, granular shear flows in a two-dimensional Couette geometry~\cite{Hsiau02,Utter04}.

One of the key features of flows of polydisperse particles is segregation~\cite{Andreotti13}. This is likely one of the most relevant problems in granular mixtures, from practical and fundamental points of view. However, despite many industrial and scientific progresses made in the past few years, the mechanisms involved in the segregation phenomenon are still poorly understood.
In particular, in the context of kinetic theory, many different papers have addressed the study of segregation~\cite{Jenkins87,Puri99,Jenkins02,Trujillo03,Gray05,G06,G08,Marks11,Gray18,Jing21}. 
On the other hand, computer simulations of bidisperse granular mixtures under USF (and without any influence of gravity)~\cite{Alam03} have not found any sign of large-scale size segregation.
Another type of works has studied segregation in flows down inclined slopes in which approximate simple shear flows have been realized, at least, in the bulk regions away from the bottom boundaries and surfaces. It is remarkable that the trigger of the segregation is the deviation from the USF of the velocity profile as has been reported in Ref.~\cite{Jing21}. This suggests that segregation can be observed even for dilute mixtures without the influence of gravity, if we drive a shear flow through a boundary. In other words, the segregation is localized near the boundaries.

Previous studies of granular binary mixtures based on the kinetic theory have mainly focused on obtaining the Navier--Stokes transport coefficients of the mixture by considering states close to the homogeneous cooling state~\cite{Garzo} and/or close to driven stationary homogeneous states \cite{Sarracino10,Khalil13,Khalil18}. The results are more scarce in the study of the rheological properties of granular binary mixtures under USF~\cite{Montanero02,Alam03,L04,Chamorro23}. 
As expected, the results show that the mixture is non-Newtonian and in some cases, the effect of bidispersity enhances the non-Newtonian character of the fluid. 
Since the USF is spatially homogeneous in the frame moving with the linear velocity field, no segregation appears in the system. However, when the USF state is slightly perturbed by small density and temperature gradients, a non-vanishing mass flux is present and the corresponding components of the diffusion tensors have been determined in the tracer limit in Refs.\ \cite{Garzo02,Garzo07d}. 
The knowledge of the shear-rate dependence of the above diffusion tensors has allowed to analyze thermal diffusion segregation induced by the presence of a temperature gradient orthogonal to the shear flow plane \cite{GV10}.

Nevertheless, so far and to the best of our knowledge, there are few studies of binary mixtures of inertial suspensions including diffusion processes, in which the rheology of inertial suspensions drastically depends on the shear rate. 
Thus, as already did in the case of granular mixtures \cite{Garzo02,Garzo07d}, one has to determine the rheological properties of sheared binary mixtures of inertial suspensions as a first step before considering the segregation problem. 
Once rheology is known, the components of the diffusion tensors can be determined by using a similar procedure as the one followed in (dry) granular mixtures. Therefore, the study of the rheology of a \emph{dilute} binary mixture of inertial suspension is an important issue.


Beyond dilute granular flows, it is quite apparent that there are many exotic rheological processes in dense flows. These processes include glass transitions, shear jamming, jamming, and DST~\cite{Barnes89,Mewis11,Otsuki11,Seto13,Brown14,Lootens05,Cwalina14,Bi11,Fall15,Peters16,Otsuki20}.
Such exotic processes cannot be observed in monodisperse systems but they can be observed only in mixtures {when} the volume fraction exceeds the transition point for crystallization of identical spheres at the volume fraction $\varphi=0.49$.

In this paper, we focus on the rheology of a dilute binary mixture under USF. As in our previous works \cite{Hayakawa17,Takada20}, the influence of the interstitial gas on solid particles is accounted for in an effective way by means of (i) a deterministic drag force proportional to the particle velocity and (ii) a stochastic Langevin-like term. 
While the first contribution attempts to model the friction of grains on the viscous fluid (a collection of gas molecules), the second term mimics the energy gained by the solid particles due to their interactions with the particles of the surrounding gas. 
The corresponding set of two coupled Boltzmann kinetic equations is solved by two complementary and independent routes: (i) Grad's moment method and (ii) event-driven simulations for hard spheres (EDLSHS). 
The comparison between kinetic theory and EDLSHS allows us to verify the reliability of the theoretical predictions as the first step to tackle the behavior of sheared binary mixtures of inertial suspensions. 
Our (approximate) analytical results of the rheological properties of the mixture (the ratio $T_1/T_2$ between the partial temperatures and the pressure tensor) agree well with simulations for conditions of practical interest. 
In particular, the temperature ratio $T_1/T_2$ and the viscosity ratio $\eta_1/\eta_2$ (where $\eta_i$ is the partial contribution of the component $i$ to the total shear viscosity $\eta=\eta_1+\eta_2$) exhibit a DST-like transition for sufficiently high values of the size ratio $\sigma^{(1)}/\sigma^{(2)}$. 
As a complement, we have also compared the velocity distribution function obtained by both Grad's moment method and a kinetic model with the one obtained by EDLSHS. 

The contents of the paper are as follows. In Sect.\ \ref{sec2}, we introduce the Langevin model and Boltzmann equation for a binary mixture of inertial suspensions under a simple shear. Section \ref{sec:rheology} deals with the theoretical procedure to derive the rheology of inertial suspensions in USF. 
Section \ref{sec:comparison} is the main part of this paper, in which we present the theoretical and numerical results and find a new rheological phase transition similar to DST. The velocity distribution function is also studied by comparing the results from Grad's approximation and simulations. 
In Sect.\ \ref{sec:conclusion}, we discuss and conclude our results. 
Moreover, there are several appendices to explain the technical details of the paper. 
In Appendix \ref{sec:Delta_T_delta_T}, the difference between $P_{yy}^{(i)}$ and $P_{zz}^{(i)}$ is discussed.
In Appendix \ref{sec:Lambda_linear}, we provide some mathematical steps to compute the collisional moment needed to determine the components of the pressure tensor from Grad's method. 
The detailed rheological properties for the temperature ratio, temperature, and viscosity are discussed in Appendix \ref{sec:detailed_rheology}.
Appendix \ref{sec:appear_disappear} discusses how the discontinuous transition appears/disappears when we change the parameters of the mixture.
The tracer limit of the theory is briefly presented in Appendix \ref{sec:tracer_limit} while Appendix \ref{sec:BGK} gives the exact solution to a Bhatnagar--Gross--Krook (BGK)-like kinetic model for granular mixtures in the high shear rate regime. 
This solution provides a two-dimensional velocity distribution function. 
Finally, the one-dimensional velocity distribution function is displayed in Appendix \ref{sec:1D_VDF} with a comparison with the one obtained from computer simulations.

\section{Basic equations for a binary mixture of inertial suspension under uniform Shear Flows}
\label{sec2}

In this section, we present the basic equations describing a dilute binary mixture of inertial suspensions under USF.
In the first subsection, we introduce the Langevin equation characterizing the motion of each particle activated by the thermal noise caused by collisions with the environmental molecules. In the second subsection, we write the corresponding set of two coupled nonlinear Boltzmann kinetic equations for the bidisperse inertial suspension in the low-density regime.

\subsection{Langevin equation}

We consider a three-dimensional binary mixture of inertial suspension modeled as a mixture of \emph{inelastic} hard spheres of masses $m_i$ and diameters $\sigma^{(i)}$ ($i=1,2$). {For the sake of simplicity, we assume that the spheres are completely smooth and hence, collisions among all pairs are characterized by (positive) constant coefficients of normal restitution} $e_{ij}\leq 1$, where the subscripts $ij$ denote the species $i$ and $j$, respectively. Let us use the notations $\bm{v}_1^{(i)}$ and $\bm{v}_2^{(j)}$ when the particle 1 (species $i$) collides with the particle 2 (species $j$).
The post-collisional velocities $\bm{v}_1^{(i)\prime}$ of particles 1 (species $i$) and $\bm{v}_2^{(j)\prime}$ for 2 (species $j$) are expressed as
\begin{align}
\begin{cases}
	\displaystyle \bm{v}_1^{(i)\prime}
	= \bm{v}_1^{(i)} -\frac{m_{ij}}{m_i} \left(1+e_{ij}\right)\left(\bm{v}_{12}^{(ij)}\cdot \widehat{\bm{\sigma}}\right)
	\widehat{\bm{\sigma}},\\
	\displaystyle \bm{v}_2^{(j)\prime}
	= \bm{v}_2^{(j)} +\frac{m_{ij}}{m_j} \left(1+e_{ij}\right)\left(\bm{v}_{12}^{(ij)}\cdot \widehat{\bm{\sigma}}\right)
	\widehat{\bm{\sigma}},
\end{cases}
	\label{eq:inelastic_collision}
\end{align}
where we have introduced the pre-collisional velocities of particles $\bm{v}_1^{(i)}$ for 1 (species $i$) and 2 (species $j$), $\bm{v}_{12}^{(ij)}\equiv \bm{v}_1^{(i)}-\bm{v}_2^{(j)}$, the unit normal vector at contact $\hat{\bm{\sigma}}$, and the reduced mass $m_{ij}\equiv m_im_j/(m_i+m_j)$.

\begin{figure}[htbp]
	\centering
	\includegraphics[width=0.5\linewidth]{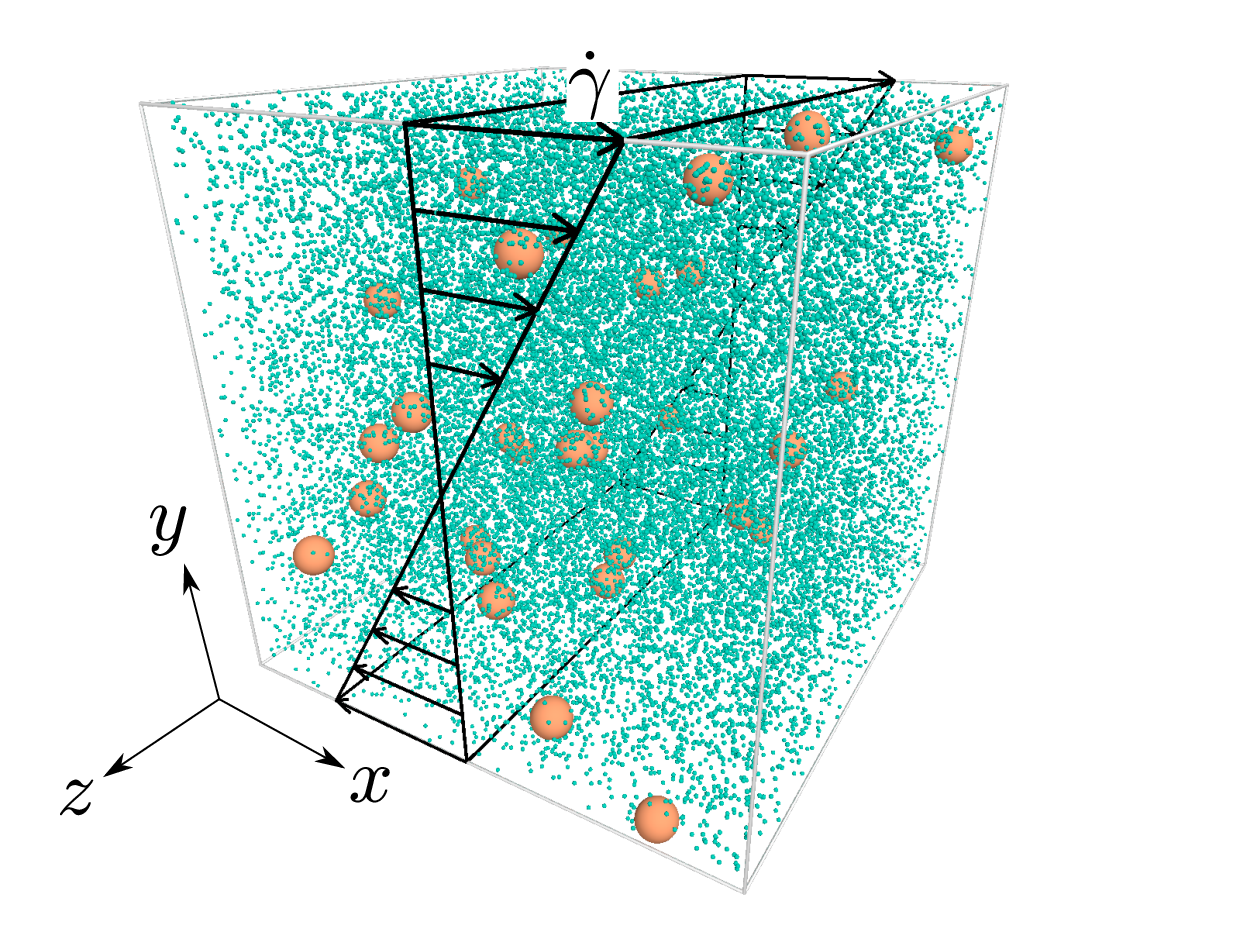}
	\caption{Setup of our system.
	Two species of particles are distributed in a fluidized inertial suspension characterized by the temperature $T_{\rm env}$.
	The shear is applied with the shear rate $\dot\gamma$ in the $xy$ plane, where the $x$ and $y$ axes are the shear direction and the velocity gradient direction, respectively.
	Here, we use $N=30000$ particles with the size and number ratio as $\sigma^{(1)}/\sigma^{(2)}=10.0$ and $N_1/N_2=30/29970=1/999$, respectively.}
	\label{fig:setup}
\end{figure}

The inertial suspension we consider is subjected to a steady simple shear flow in the $x$ direction as shown in Fig.\ \ref{fig:setup}. The equation of motion for the $k$-th particle of the species $i$ is described by the Langevin equation
\begin{equation}
	\frac{d\bm{p}_k^{(i)}}{dt} = -\zeta_i \bm{p}_k^{(i)} + \bm{F}_k^{\rm imp} + m_i \bm{\xi}^{(i)}_k,
	\label{eq:Langevin}
\end{equation}
where $\zeta_i$ is the drag coefficient acting on the particle of species $i$ from the environmental fluid, and
$\bm{p}_k^{(i)}\equiv m_i(\bm{v}_k^{(i)} - \dot\gamma y_k^{(i)} \bm{e}_x)$ is the peculiar momentum of the $k$-th particle with velocity $\bm{v}_k^{(i)}$. Here, $\dot\gamma$ and $\bm{e}_x$ are the shear rate and unit vector in the sheared ($x$) direction, respectively.
If hard-core grains are subjected to the Stokes' drag, $\zeta_i$ is simply proportional to $\sigma^{(i)}$ and $\sqrt{T_{\rm env}}$, where $T_{\rm env}$ is the environmental temperature. 
When we adopt the mean diameter $\overline{\sigma}\equiv (\sigma^{(1)}+\sigma^{(2)})/2$ and drag coefficient $\overline{\zeta}\equiv (\zeta_1+\zeta_2)/2$, the coefficient $\zeta_i$ satisfies $\zeta_i/\overline{\zeta}\propto \sigma^{(i)}/\overline{\sigma}$. For denser flows, the dependence of $\zeta_i$ on the parameters of the mixture is more complex \cite{GKG20,Gonzalez22}.   
In Eq.~\eqref{eq:Langevin}, $\bm{F}_k^{\rm imp}$ expresses the impulsive force accounting for the collisions while the noise term $\bm{\xi}^{(i)}_k(t)=\xi^{(i)}_{k,\alpha}\bm{e}_\alpha$ (the unit vector $\bm{e}_\alpha$ in the $\alpha-$direction) satisfies the fluctuation-dissipation relation \cite{K07}:
\begin{equation}
	\langle \bm{\xi}^{(i)}_k(t) \rangle = \bm{0},\quad
	\left\langle \xi^{(i)}_{k,\alpha}(t) \xi^{(j)}_{\ell,\beta} (t^\prime) \right\rangle
	= \frac{2\zeta_i T_{\rm env}}{m_i} \delta_{ij} \delta_{k\ell}\delta_{\alpha\beta} \delta(t-t^\prime).
\end{equation}

\subsection{Boltzmann equation}

If the density of {the solid particles} is low enough, the Langevin equation \eqref{eq:Langevin} can be converted into the Boltzmann kinetic equation for the distribution function $f_i(\bm{r}, \bm{v}, t)$ for the species $i$ of the dilute binary mixture of inertial suspensions. {The set of coupled Boltzmann equations read}
\begin{equation}
	\left(\frac{\partial}{\partial t}+\bm{v}\cdot \bm\nabla\right)f_i\left(\bm{r}, \bm{v}, t\right)
	=\zeta_i \frac{\partial}{\partial \bm{v}}\cdot
		\left[ \left( \bm{v} + \frac{T_{\rm env}}{m_i} \frac{\partial}{\partial \bm{v}} \right)
			f_i\left(\bm{r}, \bm{v},t\right)\right]
	+ \sum_j J_{ij}\left[\bm{v}|f_i, f_j\right],
	\label{eq:Boltzmann}
\end{equation}
with the collision integral \cite{Garzo}
\begin{align}
	J_{ij}\left[\bm{v}_1|f_i, f_j\right]
	&= \sigma_{12}^{(ij)2} \int d\bm{v}_2 \int d\widehat{\bm{\sigma}}
	\Theta\left(\widehat{\sigma}\cdot \bm{v}_{12}\right)
	\left(\widehat{\sigma}\cdot \bm{v}_{12}\right)\nonumber\\
	&\hspace{1em}\times
	\left[\frac{1}{e_{ij}^2}f_i\left(\bm{r}, \bm{v}_1^{\prime\prime},t\right)
		f_j\left(\bm{r}, \bm{v}_2^{\prime\prime},t\right)
		- f_i\left(\bm{r}, \bm{v}_1,t)f_j(\bm{r}, \bm{v}_2,t\right)\right],
	\label{eq:J}
\end{align}
where we have introduced $\sigma_{12}^{(ij)} \equiv (\sigma_1^{(i)} + \sigma_2^{(j)})/2$.

From the distribution $f_i$, one can define the number density of species $i$ as
\begin{equation}
	\label{ni}
	n_i(\bm{r},t)=\int d\mathbf{v} \; f_i(\bm{r}, \bm{v}, t),
\end{equation}
the flow velocity $\bm{U}_i$ of species $i$ as
\begin{equation}
	\label{Ui}
	U_i(\bm{r},t)=\int d\mathbf{v} \; \bm{v}\; f_i(\bm{r}, \bm{v}, t),
\end{equation}
and the partial temperature $T_i$ of species $i$
\begin{equation}
	\label{Ti}
	\frac{3}{2}n_i(\bm{r},t) T_i(\bm{r},t)=\int d\mathbf{v} \; \frac{m}{2} V(\bm{r},t)^2\; f_i(\bm{r}, \bm{v}, t).
\end{equation}
Here, $\bm{V}(\bm{r},t)\equiv \bm{v}-\bm{U}(\bm{r},t)$ is the peculiar velocity. The mean flow velocity $\bm{U}(\bm{r},t)$ and the kinetic temperature $T(\bm{r},t)$ are defined, respectively, as
\begin{equation}
	\label{U}
	\bm{U}=\rho^{-1}\sum_i\; \rho_i \bm{U}_i, \quad T=\sum_i\; \nu_i T_i,
\end{equation}
where $n\equiv n_1+n_2$ is the total number density, $\rho_i\equiv m_in_i$ is the mass density of species $i$, $\rho\equiv \rho_1+\rho_2$ is the total mass density, and $\nu_i\equiv n_i/n=N_i/N$ is the fraction of species $i$. Here, $N_i$ is the number of particles of species $i$ and $N=N_1+N_2$.

Let us consider the macroscopic velocity satisfying
\begin{equation}
	\label{USF}
	\bm{U}_1=\bm{U}_2=\bm{U}=\dot\gamma y \bm{e}_x,
\end{equation}
where $\dot\gamma$ is the constant shear rate. In terms of the peculiar velocity $\mathbf{V}$, Eq.\ \eqref{eq:Boltzmann} can be rewritten as
\begin{align}
	&\left[\frac{\partial}{\partial t}
		+\left(\bm{V}+\dot\gamma y\bm{e}_x\right)\cdot \bm\nabla
		-\dot\gamma V_y\frac{\partial}{\partial V_x}\right]f_i\left(\bm{r}, \bm{V},t\right)\nonumber\\
	&= \zeta_i \frac{\partial}{\partial \bm{V}}\cdot
		\left[ \left( \bm{V} + \frac{T_{\rm env}}{m_i} \frac{\partial}{\partial \bm{V}} \right)
			f_i\left(\bm{r}, \bm{V},t\right)\right]
	+ \sum_j J_{ij}\left[\bm{V} |f_i, f_j\right].
	\label{eq:Boltzmann_2}
\end{align}

At a macroscopic level, the USF is characterized by uniform density and temperature and the linear velocity field \eqref{USF}. In addition, at a more fundamental level, the USF is defined as that which is spatially uniform in the Lagrangian frame moving with the velocity field \eqref{USF}. In this frame, $f_i(\bm{r},\bm{v},t)=f_i(\bm{V},t)$ and Eq.\ \eqref{eq:Boltzmann_2} is reduced to the equation for the velocity distribution function:
\begin{equation}
	\left(\frac{\partial}{\partial t}
		-\dot\gamma V_y \frac{\partial}{\partial V_x}\right)f_i\left(\bm{V},t\right)
	=\zeta_i \frac{\partial}{\partial \bm{V}}\cdot
		\left[ \left( \bm{V} + \frac{T_{\rm env}}{m_i} \frac{\partial}{\partial \bm{V}} \right)
			f_i\left(\bm{V},t\right)\right]
	+ \sum_j J_{ij}\left[\bm{V} |f_i, f_j\right].
	\label{eq:Boltzmann_3}
\end{equation}

One of our theoretical goals is to determine the pressure tensor
\begin{equation}
	P_{\alpha\beta} = P^{(1)}_{\alpha\beta} + P^{(2)}_{\alpha\beta},
\end{equation}
where the partial pressure tensor for species $i$ is defined as
\begin{equation}
	P^{(i)}_{\alpha\beta}=\int d\bm{V} m_i V_\alpha V_\beta f_i\left(\bm{V}\right).
\end{equation}
We use the Greek and Latin characters for $\{x, y, z\}$ and the species $\{1,2\}$, respectively. The knowledge of the pressure tensor allows one to get the rheological properties of the inertial suspension. Needless to say, the flow under USF is independent of the spatial position by its definition. Therefore, we can start from Eq.\ \eqref{eq:Boltzmann_3} as the basic equation for the theoretical analysis.

\section{Rheology under uniform shear flows}\label{sec:rheology}

In this section, we present the results of rheology for a binary mixture of inertial suspension under USF obtained from the Boltzmann equation \eqref{eq:Boltzmann_3}. There are three subsections in this Section.
In the first subsection, we summarize a general framework to determine the rheology of inertial suspensions by deriving a set of equations for the partial pressure tensors by multiplying  both sides of Eq.\ \eqref{eq:Boltzmann_3} by $m_i \bm{V}\bm{V}$ and integrating over velocity. No approximations are considered in this subsection, including the moment of the collision integral \eqref{eq:J}.
In the second subsection, we focus on the steady rheology within the framework of the kinetic theory under Grad's moment method \cite{Grad49}.
In the third subsection, we explain the concrete procedure to determine the steady rheology.

\subsection{Moment equation for the pressure tensor}

As mentioned before, the set of equations for the pressure tensor of the species $i$ is obtained by multiplying by $m_i V_\alpha V_\beta$ both sides of the Boltzmann equation \eqref{eq:Boltzmann_3} and integrating over $\bm{V}$. The result is
\begin{equation}
	\frac{\partial}{\partial t}P^{(i)}_{\alpha\beta}
	+\dot\gamma\left(\delta_{\alpha x}P^{(i)}_{y \beta} + \delta_{\beta x}P^{(i)}_{y\alpha}\right)
	= -2\zeta_i \left(P^{(i)}_{\alpha\beta} - n_i T_{\rm env} \delta_{\alpha\beta}\right)
	-\sum_{j=1}^2 \Lambda^{(ij)}_{\alpha\beta}, \quad i=1,2,
	\label{eq:P_eq}
\end{equation}
where ${\sf \Lambda}^{(ij)}$ is the collisional moment
\begin{align}
	\Lambda^{(ij)}_{\alpha\beta}
	&\equiv -\int d\bm{V} m_i V_\alpha V_\beta J_{ij}\left[\bm{V}|f_i, f_j\right].
	\label{eq:Lambda_ij}
\end{align}

Let us introduce the anisotropic temperatures
\begin{equation}
	\Delta T_i \equiv \frac{P^{(i)}_{xx} - P^{(i)}_{yy}}{n_i},\quad
	\Delta T \equiv \frac{P_{xx} - P_{yy}}{n}.
\end{equation}
It should be noted that there are some other anisotropic temperatures such as
$\delta T \equiv (P_{xx}-P_{zz})/n$, which differ from $\Delta T$ in general.
Nevertheless, we ignore the difference between $\Delta T$ and $\delta T$ in this paper, because
(i) the detection of the difference between $\Delta T$ and $\delta T$ is difficult \cite{DST16},
and (ii) the linear approximation of Grad's method used later  yields $P_{yy}^{(i)}=P_{zz}^{(i)}$ and so, $\Delta T=\delta T$.
In general, $\delta T$ differs from $\Delta T$ even for dilute systems (Refs.~\cite{Sangani96,Chamorro15}), although the previous studies confirmed that the effect of $\delta T\neq \Delta T$ is small \cite{DST16,Chamorro15}.
We want also to indicate that the difference between $\Delta T$ and $\delta T$ is almost imperceptible in the simulations (see Appendix \ref{sec:Delta_T_delta_T}) despite the requirement of long and tedious calculations for evaluating this difference \cite{DST16}. 
Therefore, for simplicity, we ignore the difference between $\Delta T$ and $\delta T$ in this paper.

If we adopt such a simplification, the evolution equations for $T_i$, $\Delta T_i$, and $P^{(i)}_{xy}$ are given by
\begin{subequations}
	\label{eq:evol_eq_dimensional}
\begin{align}
	\frac{\partial}{\partial t}T_i
	&= -\frac{2}{3n_i}\dot\gamma P^{(i)}_{xy} + 2\zeta_i \left(T_{\rm env}-T_i\right)
		-\frac{1}{3n_i}\sum_{j=1}^2 \Lambda^{(ij)}_{\alpha\alpha},\\
	\frac{\partial}{\partial t}\Delta T_i
	&= -\frac{2}{n_i}\dot\gamma P^{(i)}_{xy} - 2\zeta_i \Delta T_i
		-\frac{1}{n_i}\sum_{j=1}^2 \left(\Lambda^{(ij)}_{xx} - \Lambda^{(ij)}_{yy}\right),\\
	\frac{\partial}{\partial t}P^{(i)}_{xy}
	&= -\dot\gamma n_i \left(T_i - \frac{1}{3}\Delta T_i\right)
		-2\zeta_i P^{(i)}_{xy} - \sum_{j=1}^2 \Lambda^{(ij)}_{xy},
\end{align}
\end{subequations}
where we have introduced the environmental temperature $T_{\rm env}$ of the suspension liquid.
Note that the diagonal elements of the pressure tensor in dilute systems can be written as
\begin{align}
\label{new1}
	P^{(i)}_{xx} &= n_i \left(T_i + \frac{2}{3}\Delta T_i\right),\\
	P^{(i)}_{yy} &= n_i \left(T_i - \frac{1}{3}\Delta T_i\right).
\end{align}
In this paper, we adopt Einstein's rule for the summation, i.e., $P^{(i)}_{\alpha\alpha} = \sum_{\alpha=1}^3 P^{(i)}_{\alpha\alpha}$. {Upon deriving Eqs.\ \eqref{new1}, we recall that we have made use of the identity $P_{yy}^{(i)}=P_{zz}^{(i)}$}.

\subsection{Kinetic theory of rheology for a dilute binary mixture of inertial suspension via Grad's method}

\subsubsection{Grad's moment method for the velocity distribution function}

To determine $P^{(i)}_{\alpha\beta}$, we need to know the explicit form of the collisional moments $\Lambda^{(ij)}_{\alpha\beta}$.
This requires the knowledge of the velocity distribution functions $f_i$, which is an intricate problem even for the elastic case.
As for monodisperse inertial suspensions, a useful way to estimate ${\sf \Lambda}^{(ij)}$ is to adopt Grad's moment method \cite{Grad49} in which the true $f_i$ is approximated {by} the trial Grad's distribution~\cite{BGK2016,Hayakawa17,Takada20,Chamorro15,Grad49,Garzo02,Santos04,Garzo13}:
\begin{equation}
	f_i(\bm{V})
	\approx f_{i,{\rm M}}(\bm{V})
	\left(1+\frac{m_i}{2T_i} \Pi^{(i)}_{\alpha\beta} V_\alpha V_\beta\right),
	\label{eq:Grad}
\end{equation}
where
\begin{equation}
	\Pi^{(i)}_{\alpha\beta}
	= \frac{P^{(i)}_{\alpha\beta}}{n_i T_i}-\delta_{\alpha\beta},
\end{equation}
and $f_{i,{\rm M}}(\bm{V})$ is the Maxwellian distribution at the temperature $T_i$ of the species $i$, i.e.,
\begin{equation}
	f_{i, {\rm M}}(\bm{V})=n_i \left(\frac{m_i}{2\pi T_i}\right)^{3/2} \exp\left(-\frac{m_i V^2}{2T_i}\right).
\end{equation}
Note that in Eq.\ \eqref{eq:Grad} we have taken into account that the mass and heat fluxes of a binary mixture vanish in the USF state.

With the use of the distribution \eqref{eq:Grad}, the integrals appearing in the expression of the collisional moments ${\sf \Lambda}^{(ij)}$ can be explicitly computed.
After a lengthy calculation (see Appendix \ref{sec:Lambda_linear} for the derivation), one gets
\begin{align}
	\Lambda^{(ij)}_{\alpha\beta}
	&= \frac{2\sqrt{\pi}}{3} n_i n_j {m_{ij}} \sigma_{12}^{(ij)2} v_{\rm T}^3
		\left(\frac{\epsilon_i + \epsilon_j}{\epsilon_i \epsilon_j}\right)^{3/2} (1+e_{ij})
		\left\{\left[\lambda_{ij}-\frac{1}{2}\frac{m_{ij}}{m_i}(1+e_{ij})\right]\delta_{\alpha\beta} \right.\nonumber\\
	&\hspace{1em}\left. +2\frac{\epsilon_i \epsilon_j}{(\epsilon_i+\epsilon_j)^2}
		\left[\left(1+\frac{3}{5}\frac{\epsilon_i+\epsilon_j}{\epsilon_i}\lambda_{ij}\right)\Pi^{(i)}_{\alpha\beta}
		 -\left(1-\frac{3}{5}\frac{\epsilon_i+\epsilon_j}{\epsilon_j}\lambda_{ij}\right)\Pi^{(j)}_{\alpha\beta}\right]\right\},\label{eq:Lambda_ij_alphabeta}
\end{align}
with
\begin{align}
	\lambda_{ij}
	&\equiv 2\frac{m_i\epsilon_j - m_j\epsilon_i}{(m_i+m_j)(\epsilon_i+\epsilon_j)}
	+\frac{1}{2}\frac{m_{ij}}{m_i}(3-e_{ij}).
\end{align}
Here, we have introduced $\epsilon_i \equiv m_i T/(\overline{m} T_i)$, and the thermal velocity $v_{\rm T} \equiv \sqrt{2T/\overline{m}}$ with the mean mass $\overline{m}$ defined as $\overline{m} \equiv (m_1+m_2)/2$. 
Upon deriving Eq.\ \eqref{eq:Lambda_ij_alphabeta}, nonlinear terms in the traceless stress tensor $\Pi^{(i)}_{\alpha\beta}$ have been neglected (linear Grad's approximation). The expression \eqref{eq:Lambda_ij_alphabeta} agrees with a previous derivation of the collision integral $\Lambda^{(ij)}_{\alpha\beta}$ \cite{Montanero02}.

Now, let us rewrite the set of equations \eqref{eq:evol_eq_dimensional} in dimensionless forms.
We introduce the partial dimensionless temperatures $\theta_i$ and the anisotropic temperatures $\Delta\theta_i$ for species $i$  as
\begin{equation}
	\theta_i \equiv \frac{T_i}{T_{\rm env}},\quad
	\Delta \theta_i \equiv \frac{\Delta T_i}{T_{\rm env}}.
\end{equation}
We also introduce {the global quantities} $\theta\equiv \sum_{i=1}^2\nu_i\theta_i$ and $\Delta \theta\equiv \sum_{i=1}^2\nu_i\Delta \theta_i$, where we recall that $\nu_i = n_i/n$.

Then, the dimensionless collisional moment 
\begin{equation}
\Lambda_{\alpha\beta}^{(ij)*}\equiv\frac{\Lambda^{(ij)}_{\alpha\beta}}{n_i \overline{\sigma}^{-1}\sqrt{T_{\rm env}^3/\overline{m}}}
\end{equation}
becomes
\begin{align}
	\Lambda^{(ij)*}_{\alpha\beta}
	&= C_{ij}\theta^{3/2} \left(\frac{\epsilon_i + \epsilon_j}{\epsilon_i  \epsilon_j}\right)^{3/2}
		\left\{\left[\lambda_{ij}-\frac{1}{2}\frac{m_{ij}}{m_i}(1+e_{ij})\right]
		\delta_{\alpha\beta} \right.\nonumber\\
	&\hspace{1em}\left. 
	+2\frac{\epsilon_i \epsilon_j}{(\epsilon_i+\epsilon_j)^2}
		\left[\left(1+\frac{3}{5}\frac{\epsilon_i+\epsilon_j}{\epsilon_i}\lambda_{ij}\right) \Pi^{(i)}_{\alpha\beta}
		-\left(1-\frac{3}{5}\frac{\epsilon_i+\epsilon_j}{\epsilon_j}\lambda_{ij}\right)\Pi^{(j)}_{\alpha\beta}\right]\right\},
	\label{eq:Lambda_ij_alphabeta*}
\end{align}
with
\begin{equation}
	C_{ij}
	\equiv 8\sqrt{\frac{2}{\pi}} \frac{\nu_j}{\nu_1 \sigma^{(1)*3}+\nu_2 \sigma^{(2)*3}}
		m_{ij}^* \sigma^{(ij)*2} \varphi (1+e_{ij}),
	\label{3_20}
\end{equation}
where we have introduced the packing fraction
\begin{align}
	\varphi &\equiv
	\frac{\pi}{6}n (\nu_1 \sigma^{(1)3} + \nu_2 \sigma^{(2)3}).
\end{align}
In addition, the dimensionless reduced mass $m_{ij}^* \equiv m_i^* m_j^*/(m_i^* + m_j^*)$, $m_i^*\equiv m_i / \overline{m}$, $\sigma^{(i)*}\equiv \sigma_i/\overline{\sigma}$, $\sigma^{(ij)*}\equiv \sigma^{(ij)}/\overline{\sigma}$, and
\begin{equation}
	\xi_{\rm env}\equiv \sqrt{\frac{T_{\rm env}}{\overline{m}}}\frac{1}{\overline{\sigma}\overline{\zeta}}
    \label{eq:xi_env_def}
\end{equation}
characterizes the magnitude of the noise \cite{DST16}. In terms of the temperature ratio
\begin{equation}
	\vartheta \equiv \frac{T_1}{T_2},
	\label{eq:def_vartheta}
\end{equation}
the partial temperatures $T_1$ and $T_2$ can be written as
\begin{align}
	\frac{T_1}{T} &= \frac{\vartheta}{\nu_2 + \nu_1 \vartheta},\quad
	\frac{T_2}{T} = \frac{1}{\nu_2 + \nu_1 \vartheta},\\
	\theta_1 &= \frac{T_1}{T}\frac{T}{T_{\rm env}}
	= \frac{\vartheta\theta}{\nu_2 + \nu_1 \vartheta},\quad
	\theta_2 = \frac{\theta}{\nu_2 + \nu_1 \vartheta}.
\end{align}
Therefore, the parameters $\epsilon_i$ ($i=1,2$) can be expressed as
\begin{equation}
	\epsilon_1 = \frac{m_i}{\overline{m}}\frac{\theta}{\theta_1}
	=\frac{m_1}{\overline{m}}\frac{\nu_2+\nu_1\vartheta}{\vartheta},\quad
	\epsilon_2 =\frac{m_2}{\overline{m}}(\nu_2+\nu_1\vartheta).
\end{equation}
Using these variables, we rewrite the set of equations \eqref{eq:evol_eq_dimensional} as
\begin{subequations}\label{eq:evol_eq}
\begin{align}
	\frac{\partial}{\partial \tau} \theta_i
	&= -\frac{2}{3}\dot\gamma^* \theta_i \Pi^{(i)}_{xy} + 2\zeta_i^*\left(1-\theta_i\right)
		-\frac{1}{3}\sum_{j=1}^2 \Lambda^{(ij)*}_{\alpha\alpha},\\
	\frac{\partial}{\partial \tau}\Delta \theta_i
	&= -2\dot\gamma^* \theta_i \Pi^{(i)}_{xy} - 2\zeta_i^* \Delta \theta_i
		-\sum_{j=1}^2 \left(\Lambda^{(ij)*}_{xx} - \Lambda^{(ij)*}_{yy}\right),\\
	\frac{\partial}{\partial \tau} \left(\theta_i \Pi^{(i)}_{xy}\right)
	&= -\dot\gamma^* \left(\theta_i - \frac{1}{3}\Delta \theta_i\right)
		-2 \zeta_i^*\theta_i \Pi^{(i)}_{xy} - \sum_{j=1}^2 \Lambda^{(ij)*}_{xy},
\end{align}
\end{subequations}
where we have introduced {the scaled time $\tau\equiv t\sqrt{T_{\rm env}/\overline{m}}/\overline{\sigma}$}, the dimensionless shear rate $\dot\gamma^* \equiv \dot\gamma\overline{\sigma}\sqrt{\overline{m}/T_{\rm env}}=\dot{\gamma}/(\overline{\zeta}\xi_{\rm env})$, and the dimensionless drag coefficient $\zeta_i^* \equiv \zeta_i\overline{\sigma}\sqrt{\overline{m}/T_{\rm env}}=\zeta_i/(\overline{\zeta}\xi_{\rm env})$.
For the sake of convenience, some explicit forms of $\Lambda^{(ij)*}_{\alpha\beta}$ in Eq.\ \eqref{eq:Lambda_ij_alphabeta} can be rewritten as
\begin{subequations}
\label{eq:Lambda}
\begin{align}
	\Lambda^{(ij)*}_{\alpha\alpha}
	&= 3 C_{ij}\theta^{3/2} \widetilde{\Lambda}_{\alpha\alpha}^{(ij)},\\
	\Lambda^{(ij)*}_{xx} - \Lambda^{(ij)*}_{yy}
	&= 2C_{ij} \theta^{3/2}
		\left[\widetilde{\Lambda}_{xy}^{(ij)} \Delta\theta_i
		- \widetilde{\Lambda}_{xy}^{\prime(ij)} \Delta\theta_j\right],\\
	\Lambda^{(ij)*}_{xy}
	&= 2C_{ij} \theta^{3/2}
		\left[\widetilde{\Lambda}_{xy}^{(ij)} \theta_i \Pi^{(i)}_{xy}
		- \widetilde{\Lambda}_{xy}^{\prime (ij)} \theta_j \Pi^{(j)}_{xy}\right],
\end{align}
\end{subequations}
where
\begin{subequations}
\begin{align}
	\widetilde{\Lambda}_{\alpha\alpha}^{(ij)}
	&\equiv \left(\frac{\epsilon_i+\epsilon_j}{\epsilon_i\epsilon_j}\right)^{3/2}
		\left[2\frac{m_i^*\epsilon_j - m_j^*\epsilon_i}{(m_i^*+m_j^*)(\epsilon_i+\epsilon_j)}
		+\frac{m_{ij}^*}{m_i^*}(1-e_{ij})\right],\\
	\widetilde{\Lambda}_{xy}^{(ij)}
	&\equiv \frac{\theta_i^{-1}}{\sqrt{\epsilon_i \epsilon_j (\epsilon_i + \epsilon_j)}}
		\left(1 + \frac{3}{5}\frac{\epsilon_i + \epsilon_j}{\epsilon_i} \lambda_{ij}\right),\\
	\widetilde{\Lambda}_{xy}^{\prime (ij)}
	&\equiv \frac{\theta_j^{-1}}{\sqrt{\epsilon_i \epsilon_j (\epsilon_i + \epsilon_j)}}
		\left(1 - \frac{3}{5}\frac{\epsilon_i + \epsilon_j}{\epsilon_j} \lambda_{ij}\right).
\end{align}
\end{subequations}

\subsection{Theoretical expressions in the steady rheology}

Although the set of Eqs.\ \eqref{eq:evol_eq} applies for time-dependent states, we are mainly interested in the rheology in the steady state.
Hereafter, we focus on the steady rheology.

\subsubsection{Theoretical flow curves in steady state}

Let us solve the set of Eqs.\ \eqref{eq:evol_eq} in the steady state. In this case ($\partial_\tau=0$), the left hand side of the set \eqref{eq:evol_eq} vanishes and one gets
\begin{subequations}
\begin{align}
	0 &= -\frac{2}{3}\dot\gamma^* \theta_i \Pi^{(i)}_{xy} + 2\zeta_i^*\left(1-\theta_i\right)
		- \sum_{j=1}^2 C_{ij} \widetilde{\Lambda}_{\alpha\alpha}^{(ij)} \theta^{3/2},
	\label{eq:steady_eq1}\\
	0 &= -2\dot\gamma^* \theta_i \Pi^{(i)}_{xy} - 2 \zeta_i^*\Delta \theta_i
		-2 \sum_{j=1}^2 C_{ij} \theta^{3/2}
		\left[\widetilde{\Lambda}_{xy}^{(ij)} \Delta\theta_i
		- \widetilde{\Lambda}_{xy}^{\prime(ij)} \Delta\theta_j\right],
	\label{eq:steady_eq2}\\
	0 &= -\dot\gamma^* \left(\theta_i - \frac{1}{3}\Delta \theta_i\right)
		-2 \zeta_i^*\theta_i \Pi^{(i)}_{xy}
		- 2\sum_{j=1}^2 C_{ij} \theta^{3/2}
		\left[\widetilde{\Lambda}_{xy}^{(ij)} \theta_i \Pi^{(i)}_{xy}
		- \widetilde{\Lambda}_{xy}^{\prime (ij)} \theta_j \Pi^{(j)}_{xy}\right].
	\label{eq:steady_eq3}
\end{align}\label{eq:steady_eq}
\end{subequations}
First, from Eq.\ \eqref{eq:steady_eq1}, one obtains
\begin{equation}
	\Pi^{(i)}_{xy} = \frac{3}{\dot\gamma^*\theta_i}
	\left[\zeta_i^*(1-\theta_i)
		-\frac{1}{2}\sum_{j=1}^2C_{ij}\widetilde{\Lambda}_{\alpha\alpha}^{(ij)} \theta^{3/2}\right].
	\label{eq:Pi_i_xy}
\end{equation}
Substituting Eq.\ \eqref{eq:Pi_i_xy} into Eq.\ \eqref{eq:steady_eq2}, a set of equations which determine $\Delta \theta_1$ and $\Delta \theta_2$ can be rewritten as
\begin{equation}
	F_{i1} \Delta \theta_1 + F_{i2} \Delta \theta_2 = G_i,
\end{equation}
for $i=1,2$. Here, we have introduced the quantities
\begin{subequations}
\begin{align}
	F_{11}(\theta,\vartheta)
	&\equiv \zeta_1^*
		+\left[C_{11}\left(\widetilde{\Lambda}_{xy}^{(11)}
		- \widetilde{\Lambda}_{xy}^{\prime(11)}\right)
		+ C_{12} \widetilde{\Lambda}_{xy}^{(12)}\right]\theta^{3/2},\quad
	F_{12}(\theta,\vartheta)
	\equiv -C_{12}\widetilde{\Lambda}_{xy}^{\prime (12)} \theta^{3/2},\\
	F_{22}(\theta,\vartheta)
	&\equiv \zeta_2^*
		+\left[C_{22}\left(\widetilde{\Lambda}_{xy}^{(22)}
		- \widetilde{\Lambda}_{xy}^{\prime(22)}\right)
		+ C_{21} \widetilde{\Lambda}_{xy}^{(21)}\right]\theta^{3/2},\quad
	F_{21}(\theta,\vartheta)
	\equiv -C_{21}\widetilde{\Lambda}_{xy}^{\prime (21)} \theta^{3/2},
\end{align}
\end{subequations}
and
\begin{subequations}
\begin{align}
	G_1(\theta,\vartheta)
	&\equiv -3\zeta_1^*(1-\theta_1)
		+\frac{3}{2}\left[C_{11} \widetilde{\Lambda}_{\alpha\alpha}^{(11)}
			+C_{12} \widetilde{\Lambda}_{\alpha\alpha}^{(12)}\right]\theta^{3/2},\\
	G_2(\theta,\vartheta)
	&\equiv -3\zeta_2^*(1-\theta_2)
		+\frac{3}{2}\left[C_{21} \widetilde{\Lambda}_{\alpha\alpha}^{(21)}
			+C_{22} \widetilde{\Lambda}_{\alpha\alpha}^{(22)}\right]\theta^{3/2}.
\end{align}
\end{subequations}
Then, $\Delta \theta_1$ and $\Delta \theta_2$ can be expressed as
\begin{equation}
	\Delta \theta_1(\theta,\vartheta) = \frac{G_1F_{22} - G_2F_{12}}{F_{11}F_{22} - F_{12}F_{21}},\quad
	\Delta \theta_2(\theta,\vartheta) = \frac{G_2F_{11} - G_1F_{21}}{F_{11}F_{22} - F_{12}F_{21}}.
	\label{eq:Delta_theta_i_theta_vartheta}
\end{equation}

Substituting Eqs.\ \eqref{eq:Pi_i_xy} and \eqref{eq:Delta_theta_i_theta_vartheta} into Eq.\ \eqref{eq:steady_eq3} leads to {the relationship}
\begin{equation}\label{flow_curve_relations}
{H_1(\theta,\vartheta)K_2(\theta,\vartheta) =
H_2(\theta,\vartheta)K_1(\theta,\vartheta)},
\end{equation}
where
\begin{subequations}
\begin{align}
	H_1(\theta,\vartheta)
	&\equiv -2\left(F_{11} \dot\gamma^* \theta_1 \Pi_{xy}^{(1)} + F_{12} \dot\gamma^* \theta_2 \Pi_{xy}^{(2)}\right),\\
	H_2(\theta,\vartheta)
	&\equiv -2\left(F_{21} \dot\gamma^* \theta_1 \Pi_{xy}^{(1)} + F_{22} \dot\gamma^* \theta_2 \Pi_{xy}^{(2)}\right),
\end{align}
\end{subequations}
and
\begin{equation}
	K_1(\theta,\vartheta)
	\equiv \theta_1 - \frac{1}{3}\Delta \theta_1,\quad
	K_2(\theta,\vartheta)
	\equiv \theta_2 - \frac{1}{3}\Delta \theta_2.
\end{equation}
{Equation \eqref{flow_curve_relations}} determines the relationship between the (reduced) global kinetic temperature $\theta$ and the temperature ratio $\vartheta$. {For given values of the mixture and at a given value of $\theta$, we determine the temperature ratio $\vartheta$ [defined in Eq.\ \eqref{eq:def_vartheta}] by numerically solving Eq.\ \eqref{flow_curve_relations}}.
As will be shown in the next section, we find a solution of $\theta$ by fixing $\vartheta$ in the intermediate shear regime where the size ratio becomes large by fixing the volume ratio. Once we determine {this} relationship, we can draw the flow curve with the aid of Eq.\ \eqref{flow_curve_relations}, where the shear rate is given by
\begin{equation}
	\label{eq:theta_vartheta_gammadot}
	\dot\gamma^* = \sqrt{\frac{H_1(\theta, \vartheta)}{K_1(\theta, \vartheta)}}.
\end{equation}

\section{Comparison between theory and simulation}\label{sec:comparison}

In this section, we compare the theoretical results obtained in sec.\ \ref{sec:rheology} with those of EDLSHS \cite{Scala12}. {We will consider binary mixtures constituted by species of the same mass density [$m_1/m_2=(\sigma^{(1)}/\sigma^{(2)})^3$] and a (common) coefficient of restitution $[e_{11}=e_{12}=e_{21}=e_{22}\equiv e]$}.
In the first subsection, we examine the case of an equimolar mixture ($\nu_1=\nu_2=1/2$ or $N_1=N_2$) while the general case of  $N_1\ne N_2$ will be analyzed in the second subsection. In particular, we find a new DST-like rheological phase transition for $N_1 \ll N_2$ when we fix the volume ratio, i.e., a binary mixture in which the concentration of one of the species (the large tracer particles 1) is much more smaller than that of the other species (the small particles 2).

\subsection{The rheology for $N_1=N_2$}
\begin{figure}[htbp]
	\centering
	\includegraphics[width=\linewidth]{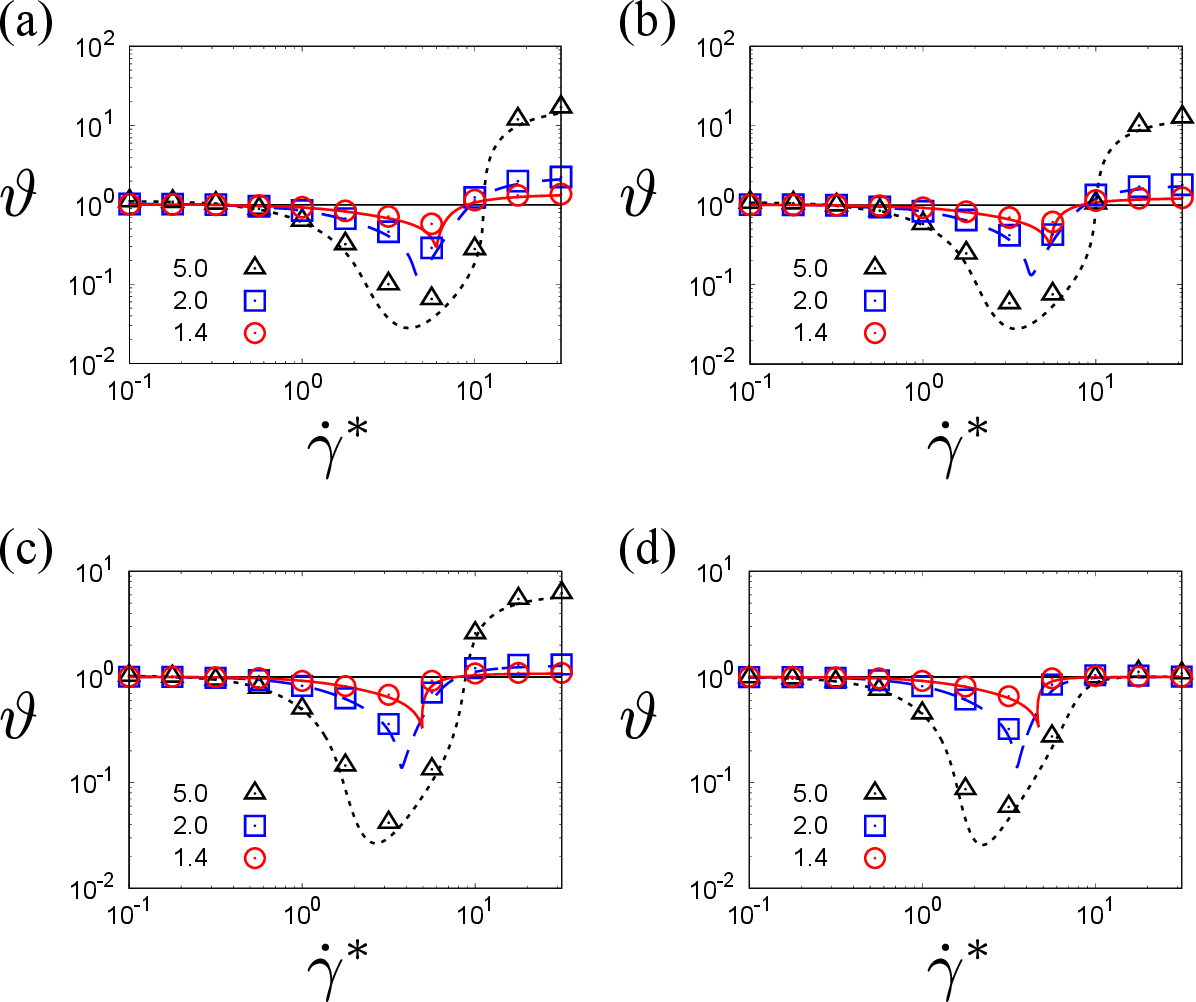}
	\caption{Temperature ratio $\vartheta=T_1/T_2$ against the dimensionless shear rate $\dot\gamma^*$ for $\sigma^{(1)}/\sigma^{(2)}=1.4$ (solid line and open circles), $2.0$ (dashed line and open squares), and $5.0$ (dotted line and open triangles) when we fix $\varphi=0.01$, $\xi_{\rm env}=1.0$, and $\nu_1=\nu_2=1/2$ for (a) $e=0.5$, (b) $0.7$, (c) $0.9$, and (d) $1$.
	The lines and symbols correspond to the steady theoretical solutions \eqref{eq:theta_vartheta_gammadot} and the simulation results ($N=1000$), respectively.}
	\label{fig:vartheta_shear}
\end{figure}
\begin{figure}[htbp]
	\centering
	\includegraphics[width=\linewidth]{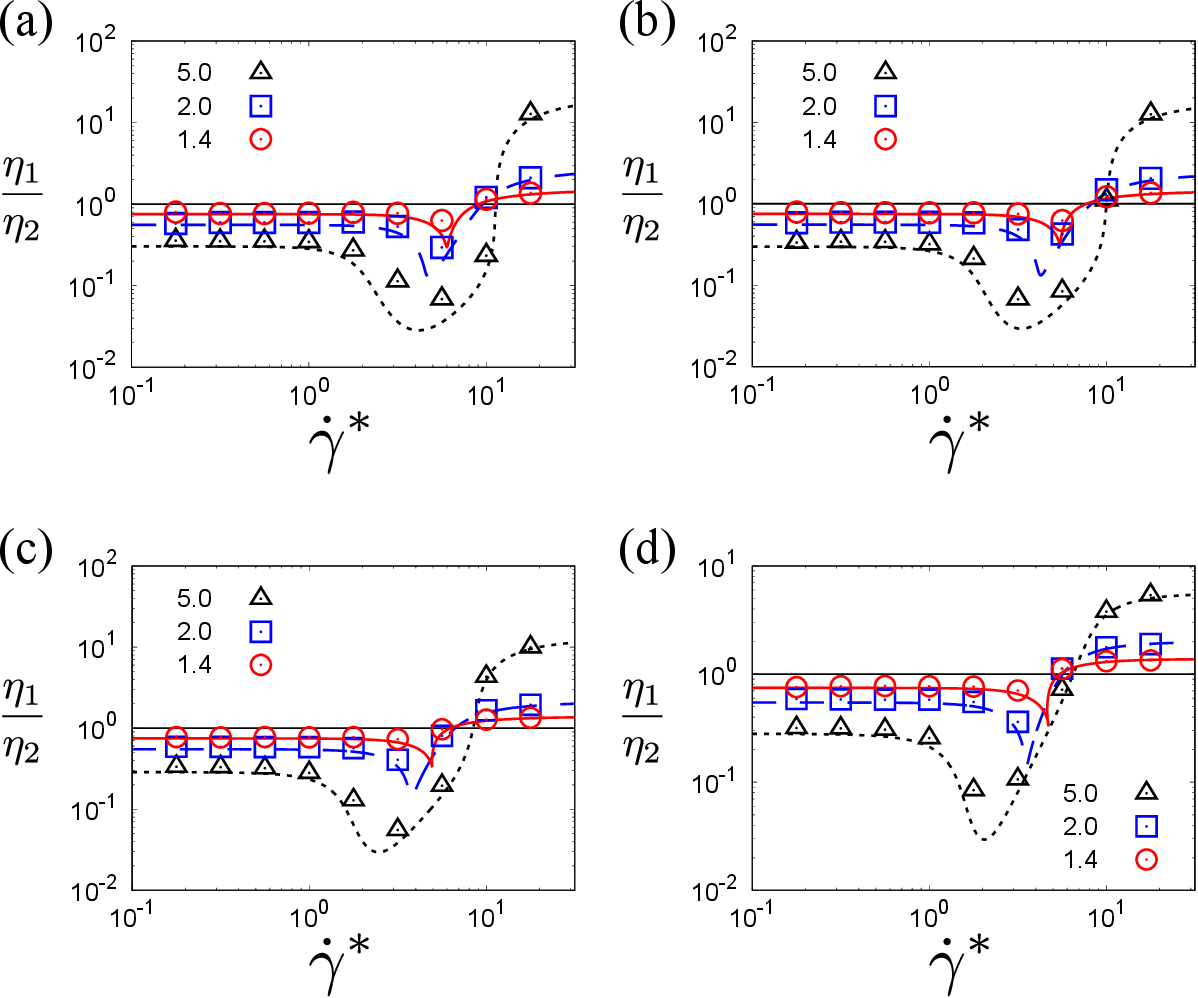}
	\caption{Viscosity ratio $\eta_1/\eta_2$ against the dimensionless shear rate $\dot\gamma^*$ for $\sigma^{(1)}/\sigma^{(2)}=1.4$ (solid line and open circles), $2.0$ (dashed line and open squares), and $5.0$ (dotted line and open triangles)  when we fix $\varphi=0.01$, $\xi_{\rm env}=1.0$, and $\nu_1=\nu_2=1/2$ for (a) $e=0.5$, (b) $0.7$, (c) $0.9$, and (d) $1$.
	The lines and symbols correspond to the steady theoretical solutions \eqref{eq:eta_i_gammadot} and the simulation results ($N=1000$), respectively.}
	\label{fig:eta_ratio_shear}
\end{figure}
In this subsection, we present the results of EDLSHS to verify the validity of the predictions of the kinetic theory for $N_1=N_2$.
In this case, we should note that the occupied volume is dominated by the large grains for a large size ratio $\sigma^{(1)}/\sigma^{(2)}$.
For example, the ratio of occupied volumes ${\mathcal{V}} \equiv N_1\sigma^{(1)3}/(N_2\sigma^{(2)3})$ becomes $125$ if we adopt $\sigma^{(1)}/\sigma^{(2)}=5.0$.
The results of EDLSHS under the control of $N_1/N_2$ with fixing $\mathcal{V}$ will be discussed in the next subsection.
For the comparison of the theoretical results with those of EDLSHS, we have used the steady solutions of Eqs.\ \eqref{eq:steady_eq} for both elastic ($e=1$) and inelastic ($e=0.5$, $0.7$, and $0.9$) cases when we fix $N=1000$, $\varphi=0.01$, $\xi_{\rm env}=1.0$, and $\nu_1=\nu_2=1/2$.

Figures \ref{fig:vartheta_shear} (for $\vartheta=T_1/T_2$) and \ref{fig:eta_ratio_shear} (for $\eta_1/\eta_2$) show some characteristic rheological flow curves for binary mixtures for both elastic and inelastic cases. Here, we have introduced the viscosity $\eta_i$ for species $i$ as
\begin{equation}
	\eta_i \equiv -P_{xy}^{(i)}/\dot\gamma.\label{eq:eta_i_gammadot}
\end{equation}

Now, let us focus on the plot of the temperature ratio $\vartheta\equiv T_1/T_2$ against the reduced shear rate $\dot\gamma^*$ in Fig.\ \ref{fig:vartheta_shear}.
In the low shear regime, the temperature ratio converges to unity as shown in Fig.\ \ref{fig:vartheta_shear}.
This is because the temperatures of both the larger and smaller particles are determined by the thermal noise of the background fluid.
On the other hand, the temperature ratio converges to a constant in the high shear regime, which is determined by the interparticle inelastic collisions. Note that this converged value agrees with the one previously obtained for granular gases \cite{Montanero02}.

Interestingly, our theory predicts the existence of a negative peak for $\vartheta$ in the intermediate shear regime.
In particular, there exists a cusp for smaller values of $\sigma^{(1)}/\sigma^{(2)}$ at a certain shear rate {$\dot\gamma_{\rm c}^*$} (at which $|\partial \vartheta/\partial \dot\gamma^*|\to \infty$; see Fig.\ \ref{fig:vartheta_shear}).
Correspondingly, the ratios of the other quantities such as $\eta_1/\eta_2$ exhibit cusps around {$\dot\gamma_{\rm c}^*$} (see Fig.\ \ref{fig:eta_ratio_shear}).
It is worth remarking that these observables exhibit common features since (i) they do not have sharp minima even near the DST-like transition point of one of two species, (ii) the deviations from unity become larger with increasing the size ratio $\sigma^{(1)}/\sigma^{(2)}$, and (iii) the ratios converge to values different from unity even in the low-shear limit.
These singularities are inherently connected with the DST-like transition {observed} (see Appendix \ref{sec:detailed_rheology}) for the global kinetic temperature $\theta$ and the shear viscosity $\eta^*=\eta_1^*+\eta_2^*$ [$\eta_i^*\equiv \eta_i/(nT_{\rm env}/\overline{\zeta})$ with Eq.~\eqref{eq:eta_i_gammadot}] because the cusps vanish as the size ratio increases. Indeed, the flow curves for $\vartheta$ and $\eta_1/\eta_2$ become smooth for $\sigma^{(1)}/\sigma^{(2)}=5.0$ (see Figs.\ \ref{fig:vartheta_shear} and \ref{fig:eta_ratio_shear}).

\begin{figure}[htbp]
	\centering
	\includegraphics[width=\linewidth]{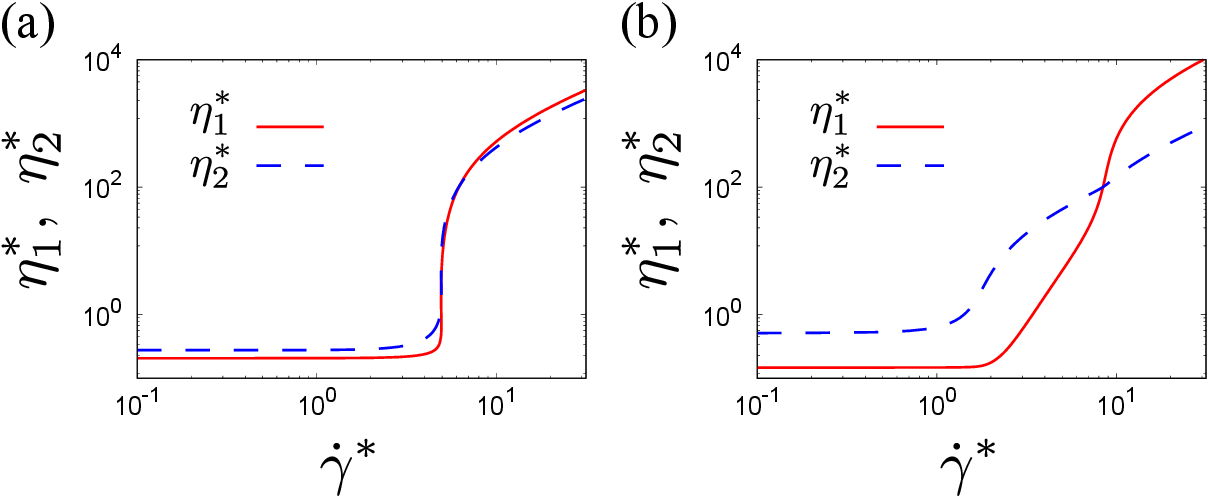}
	\caption{
	Plots of the dimensionless partial viscosities $\eta_1^*$ and $\eta_2^*$ against the dimensionless shear rate $\dot\gamma^*$ for (a) $\sigma^{(1)}/\sigma^{(2)}=1.4$ and (b) $5.0$ when we fix $\varphi=0.01$, $\xi_\mathrm{env}=1.0$, and $\nu_1=\nu_2=1/2$.}
	\label{fig:partial_visc}
\end{figure}
Let us also discuss the existence of cusps in the flow curves observed in Figs.\ \ref{fig:vartheta_shear} and \ref{fig:eta_ratio_shear} when the size ratio is small. 
As shown in Fig.~\ref{fig:partial_visc}, the partial viscosities $\eta_i^*$ also have discontinuous jumps when the mean viscosity $\eta^*=\eta_1^*+\eta_2^*$ has also this jump.
At points ($\dot\gamma_\mathrm{c}$) where $\partial\eta_i^*/\partial \dot\gamma^*\to\pm\infty$ ($i=1, 2$) are satisfied, the viscosity ratio also diverges as
\begin{equation}
    \frac{\partial}{\partial \dot\gamma^*}\left(\frac{\eta_1^*}{\eta_2^*}\right)
    = \frac{1}{\eta_2^{*2}}\left(\frac{\partial \eta_1^*}{\partial \dot\gamma^*}\eta_2^* 
    - \eta_1^* \frac{\partial \eta_2^*}{\partial \dot\gamma^*}\right)
    \to\pm\infty.
\end{equation}
This explains the reason for the existence of the cusps.

It is remarkable that the predictions of kinetic theory agree well with the simulation results of EDLSHS without any fitting parameter. Therefore, we conclude that our kinetic theory based on the Boltzmann equation with Grad's method is reliable to describe the rheology, at least, for $N_1=N_2$.

To close this subsection, we also note that the flow curves become discontinuous and continuous depending on the other parameters of the mixture.
These behaviors are discussed in Appendix \ref{sec:appear_disappear}.


\subsection{The rheology for $N_1\ne N_2$}

In this subsection, we compare the simulation results for the rheology for $N_1\ne N_2$ with those derived from the theoretical predictions by fixing the volume ratio $\mathcal{V} \equiv N_1\sigma^{(1)3}/(N_2\sigma^{(2)3})=1$. This means that the occupied volume by the large particles is the same as that by the small ones.
From the definition of the volume ratio, the size ratio correspondingly becomes $\sigma^{(1)}/\sigma^{(2)} = (N_2/N_1)^{1/3}=[(1-\nu_1)/\nu_1]^{1/3}$.
Thus, as the size ratio increases, the number of collisions between large particles decreases.
On the other hand, the impulse from the larger particle at each collision increases as compared with that from the smaller particle.

\begin{figure}[htbp]
	\centering
	\includegraphics[width=\linewidth]{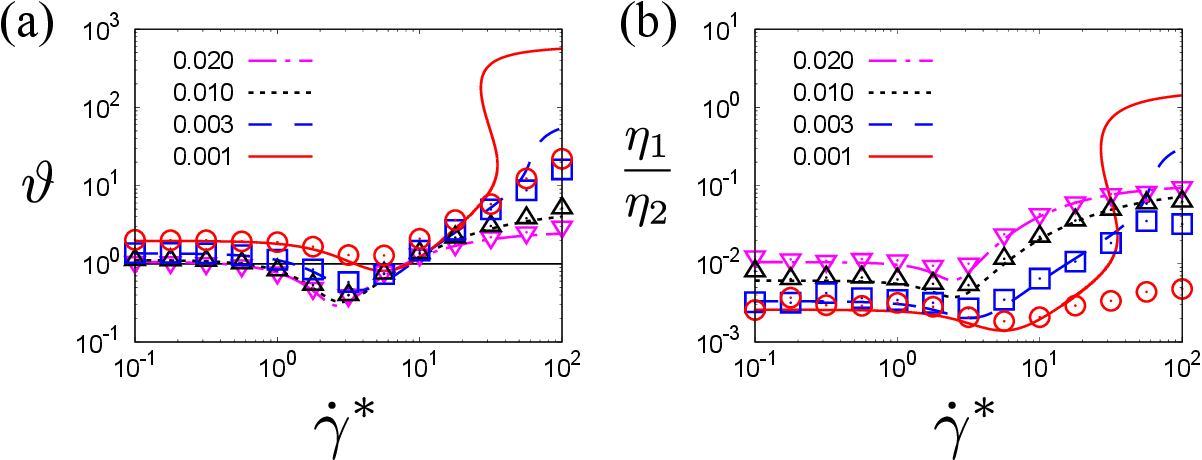}
	\caption{(a) Temperature ratio $\vartheta$ against the dimensionless shear rate $\dot\gamma^*$ for $\nu_1=1.0\times10^{-3}$ (solid line and open circles), $3.0\times10^{-3}$ (dashed line and open squares), $1.0\times10^{-2}$ (dotted line and open upper triangles), and $2.0\times10^{-2}$ (dot--dashed line and open lower triangles) when we fix $\varphi=0.01$, $\xi_{\rm env}=1.0$, $e=0.9$, and $\mathcal{V}=1$.
	(b) Viscosity ratio $\eta_1/\eta_2$ against the dimensionless shear rate $\dot\gamma$ for the same set of parameters.
	All results of simulations are obtained for $N=1000$.}
	\label{fig:vol_ratio0.50}
\end{figure}

Figures \ref{fig:vol_ratio0.50}(a) and \ref{fig:vol_ratio0.50}(b) plot the results of $\vartheta$ and $\eta_1/\eta_2$, respectively,  against $\dot\gamma^*$ for $\nu_1=1.0\times10^{-3}$ (solid line and open circles), $3.0\times10^{-3}$ (dashed line and open squares), $1.0\times10^{-2}$ (dotted line and open upper triangles), and $2.0\times 10^{-2}$ (dot--dashed line and open lower triangles) by fixing $\varphi=0.01$, $e=0.9$, and $\xi_{\rm env}=1.0$.
Here, the corresponding size ratios become (a) $\sigma^{(1)}/\sigma^{(2)} = 10.0$, (b) $6.93$, (c) $4.63$, and (d) $3.66$, respectively.
It should be noted that the number of particles of EDLSHS is fixed as $N=1000$ in Fig.\ \ref{fig:vol_ratio0.50}. 
It is quite apparent that
the theory compares well with the simulation results in the wide range of the shear rate and without any fitting parameter when the size ratio is not large (or equivalently, $\nu_1\gtrsim 3.0\times10^{-3}$ in Fig.\ \ref{fig:vol_ratio0.50}).
On the other hand, {some discrepancies} between the theoretical prediction and the EDLSHS simulations are observed  in the high shear regime when the size ratio becomes sufficiently large (see the data for $\nu_1=1.0\times10^{-3}$ in Fig.\ \ref{fig:vol_ratio0.50}).
In particular, at $\dot\gamma^*\approx 30$, the theory predicts a new DST-like transition in which the flow curve becomes an S-shape; {in this region} the temperature ratio versus the shear rate becomes a multivalued function (see Fig.\ \ref{fig:vol_ratio0.50}(a)).
Here, the upper branch becomes almost $100$ times larger compared to the lower branch.
This behavior is analogous to the ignited-quenched transitions for the shear-rate dependence of both the temperature and the viscosity for the monodisperse case \cite{DST16, Hayakawa17}.
(See Appendix \ref{sec:appear_disappear} for the minimum value of the size ratio at which this transition occurs.)
The {origin of the discrepancy between theory and simulations is essentially associated with} the suppression of the collisions between the large (tracer) particles because the number of them becomes $N_1\sim \mathcal{O}(1)$ for $\sigma^{(1)}/\sigma^{(2)}\gg1$, as discussed in the following.

\begin{figure}[htbp]
	\centering
	\includegraphics[width=\linewidth]{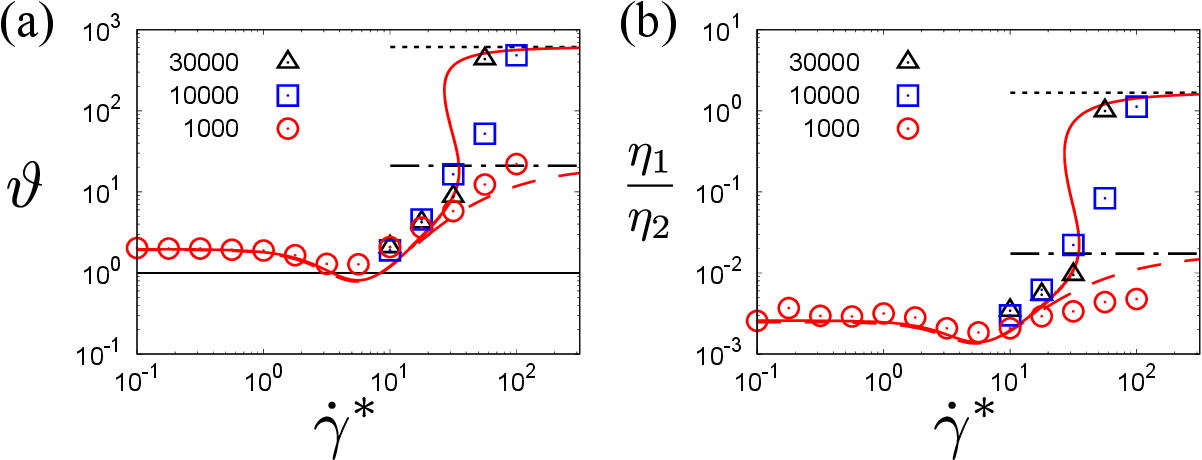}
	\caption{(a) Temperature ratio $\vartheta$ and (b) viscosity ratio $\eta_1/\eta_2$ against the dimensionless shear rate $\dot\gamma^*$ for $N=1000$ (open circles), $10000$ (open squares), and $30000$ (open triangles) when we fix $\varphi=0.01$, $\xi_{\rm env}=1.0$, $e=0.9$, $\mathcal{V}=1$, and $\nu_1=1.0\times10^{-3}$.
	The solid and dashed lines indicate the theoretical curves for Eqs.\ \eqref{eq:theta_vartheta_gammadot} and \eqref{eq:eta_i_gammadot} and the tracer limit explained in Appendix \ref{sec:tracer_limit}, respectively.
	The dotted and dot--dashed lines represent the granular gas limits for Eqs.\ \eqref{eq:theta_vartheta_gammadot} and \eqref{eq:eta_i_gammadot} and that under the tracer limit in Appendix \ref{sec:tracer_limit}, respectively.}
	\label{fig:N_dependence}
\end{figure}

To verify our conjecture, we examine the simulation results obtained for different large system sizes: $N=10000$ and $30000$.
We find that the disagreement between theory and simulation tends to decrease as the number of particles in the EDLSHS increases. {As an illustration}, Fig.\ \ref{fig:N_dependence} shows the dependence of both the temperature ratio $\vartheta$ and the viscosity ratio $\eta_1/\eta_2$ on the number of particles $N$ when we fix $\nu_1=1.0\times10^{-3}$.
Here, the relationships between the total number of particles and that of large particles correspond to $(N, N_1)=(1000, 1)$, $(10000, 10)$, and $(30000, 30)$.
As $N_1$ increases, the effect of collisions between large particles on rheology becomes non-negligible.
The collisions between large particles affect the flow curve in particular in the high shear regime.
Correspondingly, the quantities discontinuously change at a certain shear rate; this shear rate depends on the number of particles.
The above results suggest that (i) the discontinuous change predicted by the kinetic theory can be universally observed in the thermodynamic limit and (ii) {the picture of an impurity enslaved to the host fluid (namely, when the tracer-tracer collisions are neglected) is insufficient to capture the above discontinuous transition. The fact that the seemingly natural ``enslaved impurity'' picture breaks down for large shear rates has been also shown the responsible for the extreme violation of energy equipartition in a sheared granular mixture in the tracer limit \cite{GT11,GT12}.

We can also understand this finite size effect of EDLSHS when we observe the time evolution of the temperature ratio $\vartheta$ for a {very} large system.
Figure \ref{fig:evol_vartheta} exhibits the typical evolution of $\vartheta$ for $N=30000$. {The solid line refers to the solution obtained for a binary mixture assisted by Eq.\ \eqref{flow_curve_relations} with $\nu_1=1.0\times 10^{-3}$ while the dashed line corresponds to the analytical result obtained in the Appendix \ref{sec:tracer_limit} in the tracer limit (i.e., by neglecting collisions between tracer particles and by assuming that the excess species 2 is not affected by the presence of tracer particles). We observe the transient behavior in the result of EDLSHS from the tracer limit line (dashed line) to that of the (complete) solution including collisions between large particles (solid line). It is apparent that collisions between large tracer particles do not play any role in the early stage since}
the temperature ratio measured in the simulation agrees well with the tracer limit line (see the data for $\tau\lesssim 2$ in Fig.\ \ref{fig:evol_vartheta}).
As time goes on, however, those contributions become important for the rheology of the system.
As a result, the temperature ratio measured in EDLSHS starts to increase abruptly (see the data for $\tau\simeq 3$ in Fig.\ \ref{fig:evol_vartheta}), and tends to converge to the {asymptotic theoretical} value ($\tau\gtrsim5$).

\begin{figure}[htbp]
	\centering
	\includegraphics[width=0.5\linewidth]{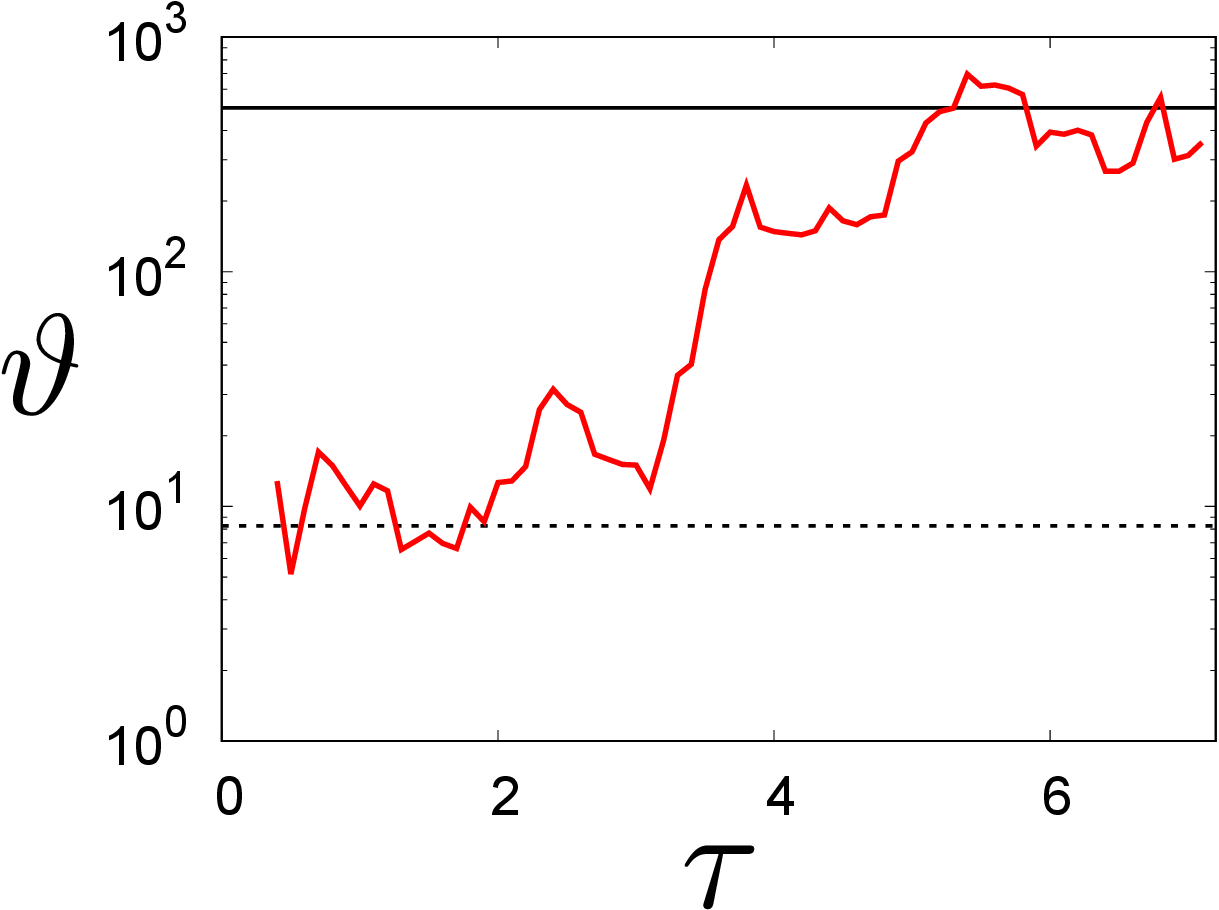}
	\caption{Time evolution of the temperature ratio $\vartheta$ when we fix $\varphi=0.01$, $\xi_{\rm env}=1.0$, $e=0.9$, $N=30000$, $\mathcal{V}=1$, $\dot\gamma^*=5.6\times10^2$, and $\nu_1=1.0\times10^{-3}$, where we have introduced the dimensionless time $\tau\equiv t\sqrt{T_{\rm env}/\overline{m}}/\overline{\sigma}$.
	The solid line {refers to the solution obtained for a binary mixture assisted by Eqs.\ \eqref{flow_curve_relations} and \eqref{eq:theta_vartheta_gammadot} with $\nu_1=1.0\times 10^{-3}$ while the dashed line corresponds to the analytical result obtained in the Appendix \ref{sec:tracer_limit} in the tracer limit}.}
	\label{fig:evol_vartheta}
\end{figure}

\begin{figure}[htbp]
	\centering
	\includegraphics[width=\linewidth]{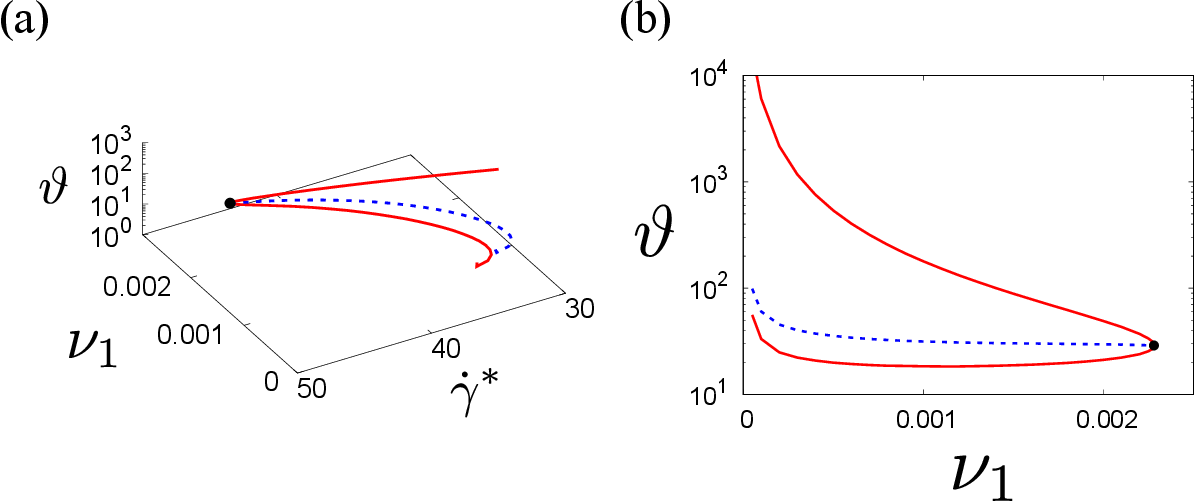}
	\caption{(a) Plot of the phase coexistence line $\partial\dot\gamma^*/\partial \vartheta=0$ (solid lines) and the spinodal line $\partial^2\dot\gamma^*/\partial \vartheta^2=0$ (dashed line) for $\varphi=0.01$, $\xi_{\rm env}=1.0$, $e=0.9$, and $\mathcal{V}=1$.
	(b) Plot of the projection of the phase coexistence line and the spinodal line onto the $(\nu_1,\vartheta)$ plane.
	The point indicates the critical point $(\nu_1, \dot\gamma^*,\vartheta)\simeq (2.28\times10^{-2}, 45.9, 28.9)$.}
	\label{fig:critical_vartheta}
\end{figure}

Let us check how the discontinuous changes of the temperature ratio and the viscosity ratio appear in the thermodynamic limit.
{According to Fig.\ \ref{fig:vol_ratio0.50}, there must exist a critical value} $\nu_{1,{\rm c}}$ of the fraction in the range $1.0\times10^{-3}<\nu_{1,{\rm c}}<3.0\times10^{-3}$.
The discontinuity is characterized by a point {(i)} at which $\partial \vartheta/\partial \dot\gamma^*\to \infty$ in the higher temperature regime and where {(ii)} the curve of $\vartheta$ versus  $\dot\gamma^*$ discontinuously changes in the lower regime.
Here, we introduce a critical temperature ratio $\vartheta_{\rm c}$, which satisfies the identities
\begin{equation}
\label{new2}
	\left(\frac{\partial \dot\gamma^*}{\partial \vartheta}\right)_{e,\varphi,\nu_1}=0,\quad
	\left(\frac{\partial^2 \dot\gamma^*}{\partial \vartheta^2}\right)_{e,\varphi,\nu_1}=0.
\end{equation}
The relations \eqref{new2} are analogous to the critical point at the first-order transition \cite{Hayakawa17}.
Figure \ref{fig:critical_vartheta} shows the {dependence of both the phase coexistence line ($\partial\dot\gamma^*/\partial \vartheta=0$) and the spinodal line ($\partial^2\dot\gamma^*/\partial \vartheta^2=0$) on the fraction fraction $\nu_1$} for $\varphi=0.01$, $\xi_{\rm env}=1.0$, $e=0.9$, and $\mathcal{V}=1$.
As expected, these lines converge to the critical values $\nu_1\simeq 2.28\times10^{-2}$, $\dot\gamma^*\simeq 45.9$, and $\vartheta\simeq 28.9$.
The finding of a DST-like rheological phase transition in the large shear rate region if the size ratio is large under fixing the volume ratio is {one of the most interesting results achieved} in this paper.

\subsection{Velocity distribution function}\label{sec:VDF}

In this subsection, let us {compare} the velocity distribution function (VDF) \eqref{eq:Grad} of Grad's moment method with the one obtained by means of simulations. {As a complement, we also include} the exact VDF of a BGK-like kinetic model in the large shear limit (see the Appendix \ref{sec:BGK}).

For later analysis, let us introduce the dimensionless velocity $\bm{c}$ and the distribution function $g_{i, {\rm G}}(\bm{c})$ as
\begin{equation}
	\bm{c}\equiv  \displaystyle\sqrt{ \frac{m_i}{2T_i} }
	\bm{V},
	\quad
	g_{i, {\rm G}}(\bm{c}) \equiv
	\displaystyle{\left(\frac{2T_i}{m_i}\right)}^{3/2}
	\frac{f_{i, {\rm G}}(\bm{V})}{n_i},
\end{equation}
where $f_{i, {\rm G}}$ stands for Grad's VDF \eqref{eq:Grad} for species $i$.
Now, we focus on the VDF of the larger particles 1. It is convenient to consider the marginal distribution $g_{1, {\rm G}}^{(xy)}$ instead of using the full three-dimensional VDF. The distribution $g_{1, {\rm G}}^{(xy)}$ is defined as
\begin{align}
	g_{1, {\rm G}}^{(xy)}(c_{x},c_{y}) &= \int_{-\infty}^\infty dc_z g_{1, {\rm G}}(\bm{c}) \nonumber\\
	&=g_{1, {\rm M}}^{(xy)}(c_{x},c_y)\Bigg[1+\Bigg(\frac{1}{2}+c_y^2-2c_x^2\Bigg)\Pi_{yy}^{(1)}+2 c_xc_y\Pi_{xy}^{(1)}\Bigg],
\label{eq:g_1xy_Grad}
\end{align}
where
\begin{equation}
	g_{1, {\rm M}}^{(xy)}(c_{x},c_y)=\frac{1}{\pi}e^{-\left(c_{x}^2+c_y^2\right)}.
\end{equation}

The VDF $g_{1, {\rm G}}^{(x,y)}(c_{x},c_y)$ in Eq.\ \eqref{eq:g_1xy_Grad} can characterize {the anisotropy of the VDF} induced by the shear flow.

\begin{figure}[htbp]
	\centering
	\includegraphics[width=\linewidth]{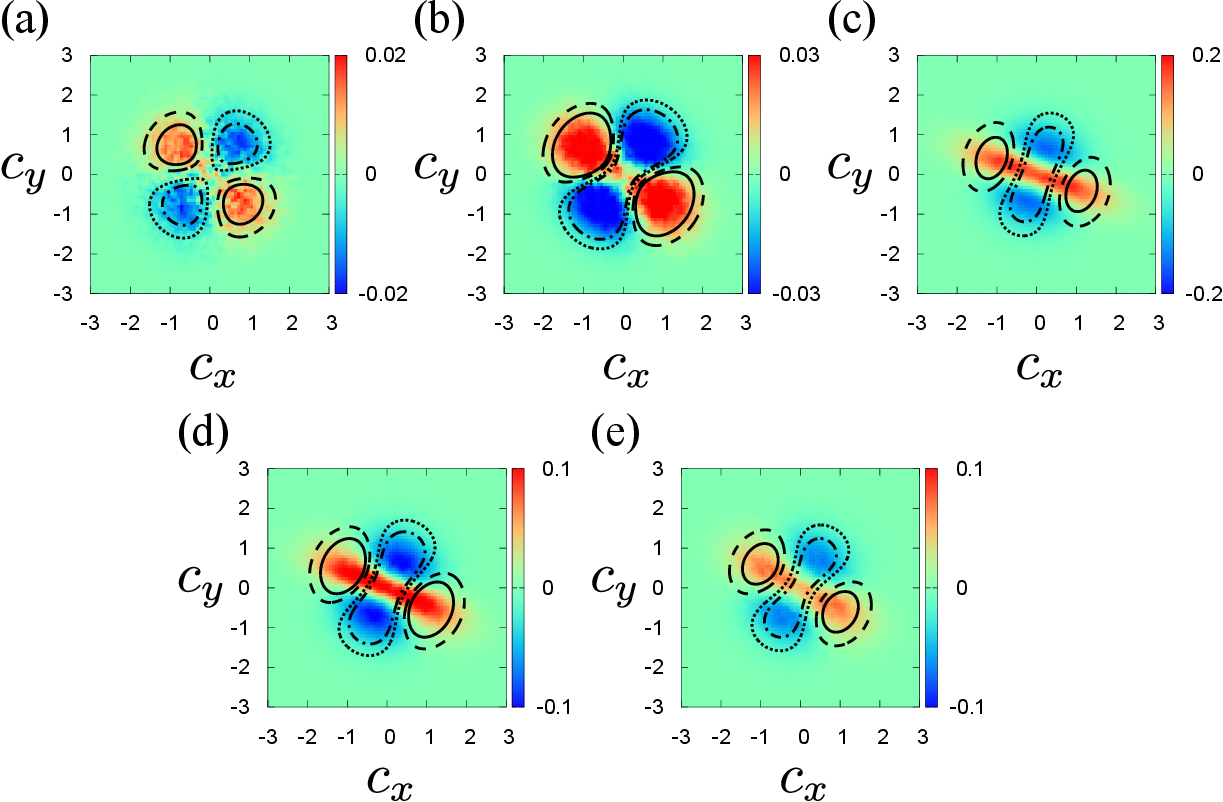}
	\caption{Velocity distribution functions of the larger particles $g_{1, {\rm G}}^{(xy)}(c_{x},c_y)-g_{1, {\rm M}}^{(xy)}(c_{x},c_y)$ in the ($c_x$, $c_y$)-plane for (a) $\dot\gamma^*=0.32$, (b) $1.0$, (c) $3.2$, (d) $10$, and (e) $32$ when we fix $\varphi=0.01$, $e=0.9$, $\xi_{\rm env}=1.0$, $\sigma^{(1)}/\sigma^{(2)}=2.0$, and $\nu_1=\nu_2=1/2$.
	The color plot corresponds to the simulation results.
	The solid, dashed, dotted, and dot--dashed lines represent the contours $0.2 c_{\rm max}$, $0.1c_{\rm max}$, $-0.1c_{\rm max}$, and $-0.2c_{\rm max}$ obtained from Grad's method \eqref{eq:g_1xy_Grad} with (a) $c_{\rm max}=0.02$, (b) $0.03$, (c) $0.2$, and (d, e) $0.1$, respectively.}
	\label{fig:prob_2.0_e0.9}
\end{figure}
\begin{figure}[htbp]
	\centering
	\includegraphics[width=\linewidth]{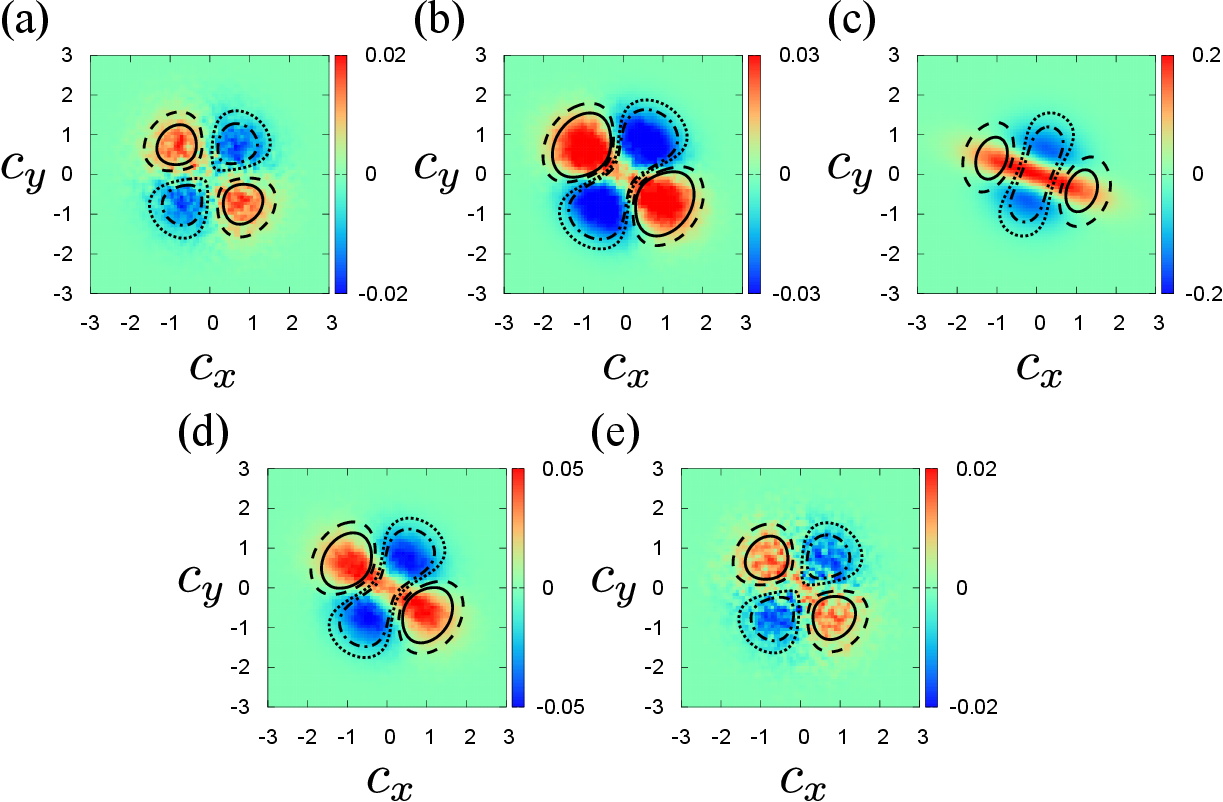}
	\caption{Velocity distribution functions of the larger particles $g_{1, {\rm G}}^{(xy)}(c_{x},c_y)-g_{1, {\rm M}}^{(xy)}(c_{x},c_y)$ in the ($c_x$, $c_y$)-plane for (a) $\dot\gamma^*=0.32$, (b) $1.0$, (c) $3.2$, (d) $10$, and (e) $32$ when we fix $\varphi=0.01$, $e=1.0$, $\xi_{\rm env}=1.0$, $\sigma^{(1)}/\sigma^{(2)}=2.0$, and $\nu_1=\nu_2=1/2$.
	The color plot corresponds to the simulation results.
	The solid, dashed, dotted, and dot--dashed lines represent the contours $0.2 c_{\rm max}$, $0.1c_{\rm max}$, $-0.1c_{\rm max}$, and $-0.2c_{\rm max}$ obtained from Grad's method \eqref{eq:g_1xy_Grad} with (a) $c_{\rm max}=0.02$, (b) $0.03$, (c) $0.2$, (d) $0.05$, and (e) $0.02$, respectively.}
	\label{fig:prob_2.0_e1}
\end{figure}
\begin{figure}[htbp]
	\centering
	\includegraphics[width=\linewidth]{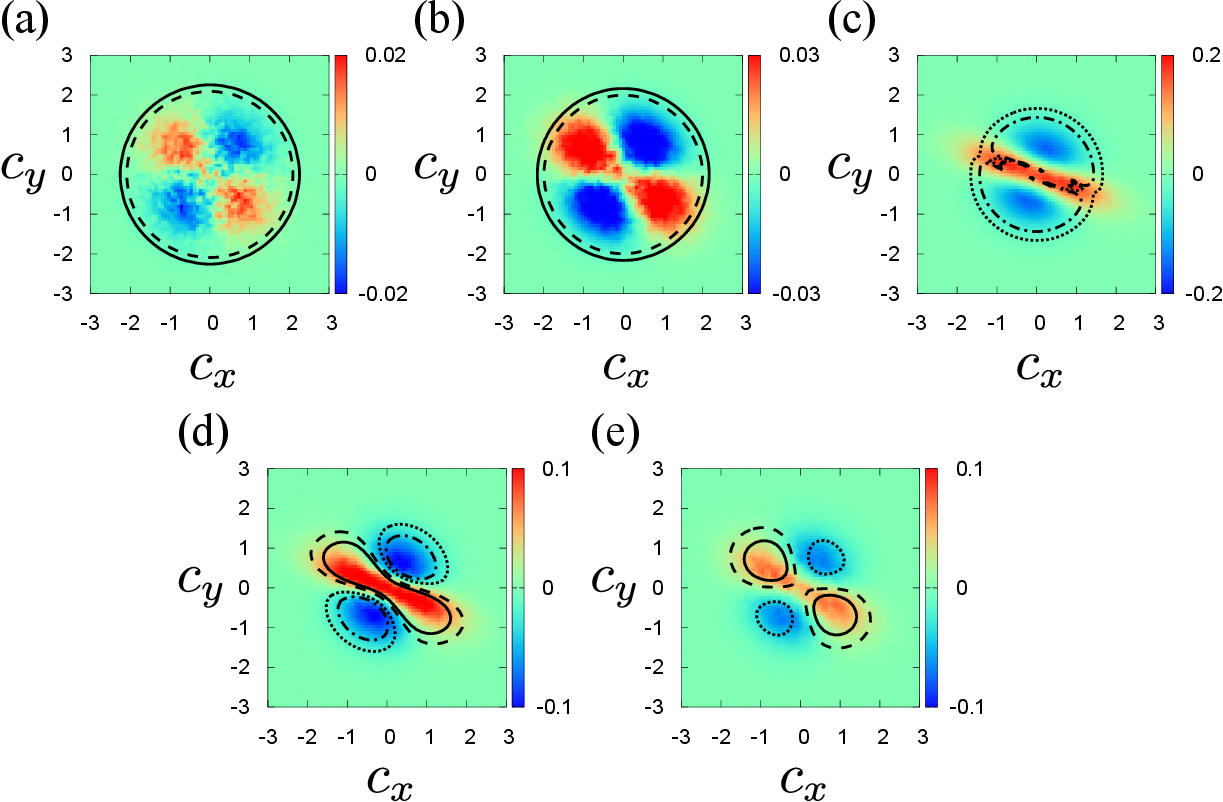}
	\caption{Velocity distribution functions of the larger particles $g_{1, {\rm G}}^{(xy)}(c_{x},c_y)-g_{1, {\rm M}}^{(xy)}(c_{x},c_y)$ in the ($c_x$, $c_y$)-plane for (a) $\dot\gamma^*=0.32$, (b) $1.0$, (c) $3.2$, (d) $10$, and (e) $32$ when we fix $\varphi=0.01$, $e=0.9$, $\xi_{\rm env}=1.0$, $\sigma^{(1)}/\sigma^{(2)}=2.0$, and $\nu_1=\nu_2=1/2$.
	The color plot corresponds to the simulation results.
	The solid, dashed, dotted, and dot--dashed lines represent the contours $0.2 c_{\rm max}$, $0.1c_{\rm max}$, $-0.1c_{\rm max}$, and $-0.2c_{\rm max}$ obtained from the BGK model \eqref{eq:g_1xy_BGK} with (a) $c_{\rm max}=0.02$, (b) $0.03$, (c) $0.2$, and (d, e) $0.1$, respectively.}
	\label{fig:prob_2.0_e0.9_BGK}
\end{figure}

Figures \ref{fig:prob_2.0_e0.9} and \ref{fig:prob_2.0_e1} present $g_{1, {\rm G}}^{(xy)}(c_x,c_y)-g_{1, {\rm M}}^{(xy)}(c_x,c_y)$ for $\dot\gamma^*=0.32$, $1.0$, $3.2$, $10$, and $32$. These values of the shear rate belong to the lower ($0.32$ and $1.0$), intermediate ($3.2$ and $10$), and higher ($32$) branches of the flow curve for $e=0.9$ in Fig.\ \ref{fig:prob_2.0_e0.9} and $e=1$ in Fig.\ \ref{fig:prob_2.0_e1}, respectively.
It is remarkable that Grad's distribution works well in the wide range of the shear rate as shown in Figs.\ \ref{fig:prob_2.0_e0.9} and \ref{fig:prob_2.0_e1}.
The corresponding trends are clearly observed when we consider the one-dimensional VDF in the $x$ direction (see the Appendix \ref{sec:1D_VDF}).
Nevertheless, it seems that the enhancement of the VDF in the shear direction is underestimated in the theoretical prediction.
It should be noted that this enhancement is suppressed for the one-dimensional VDF as shown in the Appendix \ref{sec:1D_VDF}.

We also check whether the VDF obtained from the BGK-like model can be used (see the Appendix \ref{sec:BGK} for details).
As expected, the deviation of the distribution of BGK-like model from that of the simulation is large for low shear regime. On the other hand, the agreement between BGK and simulations is reasonable in the high shear regime (see Fig.\ \ref{fig:prob_2.0_e0.9_BGK}).
In particular, the BGK distribution is more accurate than that of Grad's distribution in the small velocity region. In any case, it is important to recall that the BGK distribution obtained in the Appendix \ref{sec:BGK} only holds when $T_\text{env}=0$. This means that the possible discrepancies between the BGK distribution and simulations can be in part due to the fact that $T_\text{env}\neq 0$ in the simulations.

\section{Discussion and conclusion}
\label{sec:conclusion}

In this paper, we have theoretically derived the rheology of a dilute binary mixture of inertial suspensions under USF.
As in previous papers \cite{Hayakawa17,Takada20}, two different but complementary approaches have been employed to solve the set of coupled Boltzmann kinetic equations. On the analytical side, Grad's moment method \cite{Grad49} has been used to approximately solve the Boltzmann equation. Since the mass and heat fluxes vanish in the USF, only the partial traceless stress tensors $\Pi_{\alpha\beta}^{(i)}$ are retained in the trial distribution functions $f_i(\bm{V})$. Then, the theoretical predictions for the temperature ratio $T_1/T_2$ and the viscosity ratio $\eta_1/\eta_2$ have been compared against computer simulations based on the event-driven Langevin simulation method.
We have confirmed that the theoretical predictions agree with the results of simulation for hard spheres for various size ratios in most parameters' regions.
We have found that the temperature ratio and viscosity ratio discontinuously change at a certain shear rate as the size ratio increases. This feature cannot be captured by simulations when the size of the system is small.
The above transition is similar to DST in dense suspensions or the first-order phase transition at equilibrium.
Although the tracer limit of the theory is validated when the system size is small, the collisions between large tracer particles play dominant roles in the high shear regime.
We have also compared the velocity distribution functions obtained by Grad's method and BGK-like model with those obtained from the simulations.

There are several future perspectives. First, we plan to analyze the mass transport of impurities in a sheared inertial suspension. As already did in Ref.\ \cite{Garzo07d}, a Chapman--Enskog-like expansion around the local shear flow distribution obtained here will be considered to identify the shear-rate dependent diffusion $D_{\alpha\beta}$, pressure diffusion $D_{p,\alpha\beta}$, and thermal diffusion $D_{T,\alpha\beta}$ tensors. The determination of $D_{\alpha\beta}$, $D_{p,\alpha\beta}$, and $D_{T,\alpha\beta}$ will be discussed in a forthcoming paper. More importantly, the knowledge of the above diffusion tensors will allow us to analyze segregation by thermal diffusion \cite{GV10}. In the present paper, we have restricted to homogeneous systems, which makes the analysis easier than that for inhomogeneous systems.
However, depending on the size or density of particles, the segregation is inevitable when one considers binary mixtures.
In a sheared system, the segregation has been observed if there exists an inhomogeneous velocity profile \cite{Jing21}.
However, the velocity profile remains linear in our simulations as far as we have checked.
This linearity is violated if we consider systems under gravity or wall driven sheared systems.
We believe that this scenario of segregation can be described by a dilute system {described by} the Boltzmann equation.
We will analyze such systems in the near future.

Needless to say, we plan also to extend our analysis to moderately dense systems with the aid of the Enskog equation.
The extension is tough but straightforward using a similar procedure as the one followed for monodisperse systems \cite{Takada20}. This study will be also carried out in the future.

\section*{Acknowledgements}
S.T.\ and H.H.\ thank K.\ Saitoh and M.\ Otsuki for fruitful discussions.
The research of S.T.\ and H.H.\ was partially supported by the Grant-in-Aid of MEXT for Scientific Research (Grant No.\ JP21H01006).
The researches of S.T.\ and H.H.\ were also partially supported by the Grant-in-Aid of MEXT for Scientific Research (Grant Nos.\ JP20K14428 and JP16H04025, respectively) and ISHIZUE 2020 of the Kyoto University Research Development Program.
The research of V. G. has been supported by the Spanish Government through Grant No. PID2020-112936GB-I00 funded by MCIN/AEI/10.13039/501100011033, and from Grant IB20079 funded by Junta de Extremadura (Spain) and by ERDF A way of making Europe.
This work was initiated during a short stay of V.G. at the Yukawa Institute for Theoretical Physics (YITP, Kyoto University) supported by the YITP activity (YITP--T--18--03).  V.G.\ appreciates the warm hospitality of the YITP at that time.
S.T.\ and H.H.\ also acknowledge the warm hospitality of the Universidad de Extremadura during their stay there in 2020.

\appendix

\section{Difference between $P_{yy}^{(i)}$ and $P_{zz}^{(i)}$}\label{sec:Delta_T_delta_T}
In this Appendix, we show the difference between $P_{yy}^{(i)}$ and $P_{zz}^{(i)}$.
As mentioned in the main text, the second normal stress difference of species $i$, $N_2^{(i)}\equiv P_{yy}^{(i)}-P_{zz}^{(i)}$ is, in general, non-zero.
However, the second difference $N_2^{(i)}$ is treated as zero in the dilute limit of the kinetic theory.
Figure \ref{fig:N1N2} shows the plot of the ratio of the second to the first normal stress differences against the shear rate for $\sigma^{(1)}/\sigma^{(2)}=2.0$ and $5.0$ obtained from the simulations when we fix $\varphi=0.01$, $\xi_\mathrm{env}=1.0$, $e=0.9$, and $\nu_1=\nu_2=1/2$.
Here, we have introduced the first normal stress difference of species $i$ as $N_1^{(i)}\equiv P_{xx}^{(i)}-P_{yy}^{(i)}$.
Although the second difference $N_2^{(i)}$ is values are much smaller than the values of $N_1^{(i)}$ in the wide range of shear rates considered.
Therefore, we do not consider the difference between them in this paper.
It is noted that the second normal stress difference cannot be neglected when the volume fraction is finite.
\begin{figure}[htbp]
	\centering
	\includegraphics[width=0.5\linewidth]{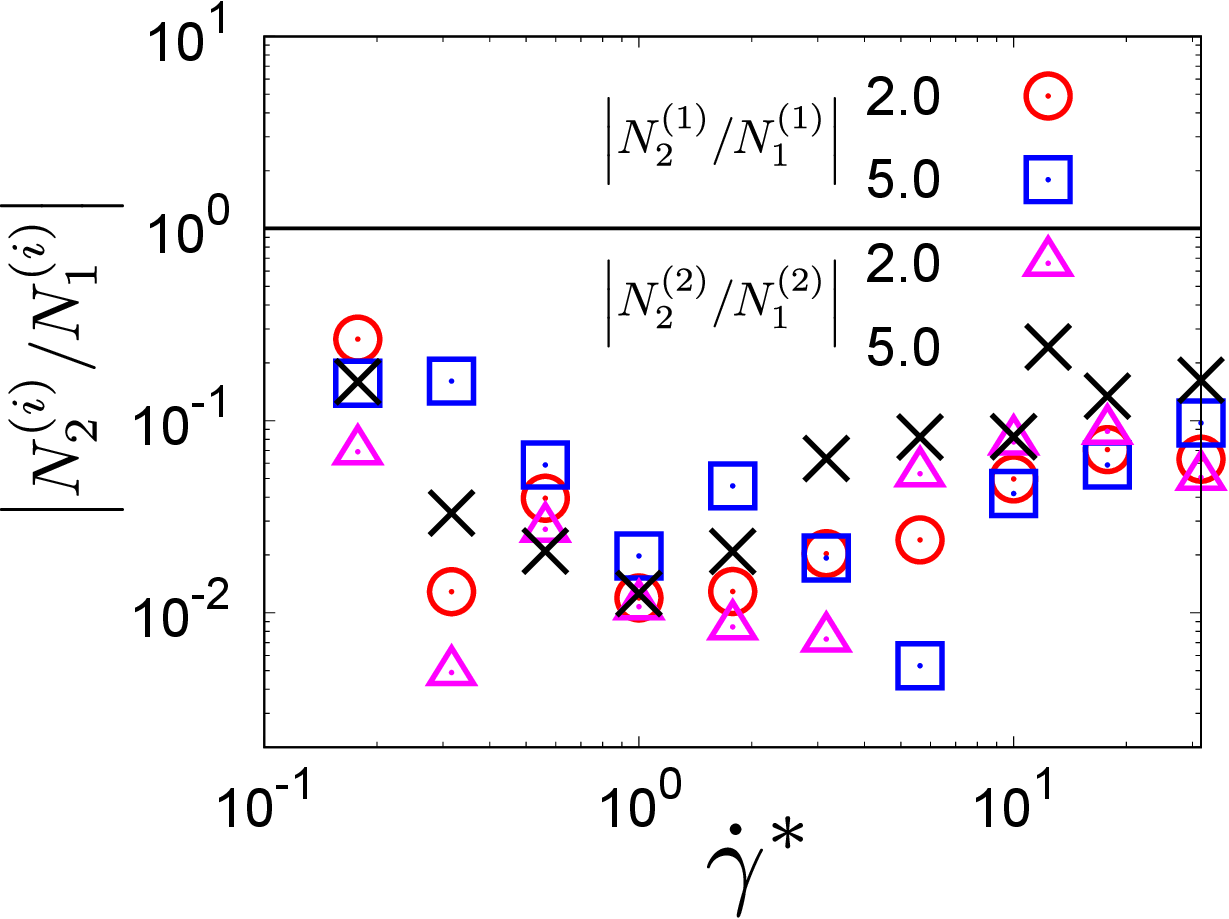}
	\caption{Plots of the ratio of the second to the first difference of each species $i$, $|N_2^{(i)}/N_1^{(i)}|$ for $\sigma^{(1)}/\sigma^{(2)}=2.0$ and $5.0$ when we fix $\varphi=0.01$, $\xi_\mathrm{env}=1.0$, $e=0.9$, and $\nu_1=\nu_2=1/2$.}
	\label{fig:N1N2}
\end{figure}

\section{Derivation of $\Lambda_{\alpha\beta}^{(ij)}$ under the linear approximation of Grad's expansion}\label{sec:Lambda_linear}

In this Appendix, we obtain the expression \eqref{eq:Lambda_ij_alphabeta} for the collisional moment $\Lambda_{\alpha\beta}^{(ij)}$.
For this purpose, we introduce the dimensionless velocities
\begin{equation}
\begin{cases}
	\displaystyle \bm{G} = \frac{m_i\bm{V}_1 + m_j\bm{V}_2}{(m_i + m_j)v_{\rm T}},\\
	\displaystyle \bm{g} =\frac{\bm{V}_1 - \bm{V}_2}{v_{\rm T}},
\end{cases}
	\label{eq:G_g}
\end{equation}
and equivalently
\begin{equation}
\begin{cases}
	\displaystyle \bm{V}_1 = \left(\bm{G} + \frac{m_{ij}}{m_i}\bm{g} \right)v_{\rm T},\\
	\displaystyle \bm{V}_2 = \left(\bm{G} - \frac{m_{ij}}{m_j}\bm{g} \right)v_{\rm T}.
\end{cases}
\end{equation}
Let us rewrite $f_i\left(\bm{V}_1\right) f_j\left(\bm{V}_2\right)$ in terms of $\bm{G}$ and $\bm{g}$.
Using {the Grad's trial distribution} \eqref{eq:Grad}, we can rewrite $f_i(\bm{V}_1)f_j(\bm{V}_2)$ as
\begin{align}
	&f_i\left(\bm{V}_1\right) f_j\left(\bm{V}_2\right)\nonumber\\
	&= n_i n_j \left(\frac{\overline{m}}{2\pi T}\right)^3 \left(\epsilon_i \epsilon_j\right)^{3/2}\nonumber\\
	&\hspace{1em}\times
	\exp\left[-(\epsilon_i + \epsilon_j)G^2
		-2\left(\frac{m_{ij}}{m_i}\epsilon_i -\frac{m_{ij}}{m_j}\epsilon_j\right)(\bm{G}\cdot \bm{g})
		-\left(\frac{m_{ij}^2}{m_i^2}\epsilon_i + \frac{m_{ij}^2}{m_j^2}\epsilon_j\right)g^2\right]\nonumber\\
	&\hspace{1em}\times
	\left[1+\epsilon_i \Pi^{(i)}_{\alpha\beta}\left(G_\alpha+\frac{m_{ij}}{m_i}g_\alpha\right)
		\left(G_\beta+\frac{m_{ij}}{m_i}g_\beta\right)
    +\epsilon_j \Pi^{(j)}_{\alpha\beta}\left(G_\alpha-\frac{m_{ij}}{m_j}g_\alpha\right)
		\left(G_\beta-\frac{m_{ij}}{m_j}g_\beta\right)\right].
	\label{eq:fi_fj}
\end{align}
We note that nonlinear contributions of the stress tensor $\Pi_{\alpha\beta}^{(i)}$ are ignored in this Appendix.
Let us rewrite the argument of the exponential part in Eq.\ \eqref{eq:fi_fj} as
\begin{align}
	&(\epsilon_i + \epsilon_j)G^2
	+2\left(\frac{m_{ij}}{m_i}\epsilon_i - \frac{m_{ij}}{m_j}\epsilon_j\right)(\bm{G}\cdot \bm{g})
		+\left(\frac{m_{ij}^2}{m_i^2}\epsilon_i + \frac{m_{ij}^2}{m_j^2}\epsilon_j\right)g^2\nonumber\\
	&= (\epsilon_i + \epsilon_j) \left[\bm{G}+\frac{m_j \epsilon_i-m_i \epsilon_j}{(m_i+m_j)(\epsilon_i+\epsilon_j)}\bm{g}\right]^2
		+ \frac{\epsilon_i \epsilon_j}{\epsilon_i + \epsilon_j}g^2.
\end{align}
Introducing $\bm{G}^\prime$ as
\begin{equation}
	\bm{G}^\prime \equiv \bm{G}+\frac{m_j \epsilon_i-m_i \epsilon_j}{(m_i+m_j)(\epsilon_i+\epsilon_j)}\bm{g},
	\label{eq:G_prime}
\end{equation}
{one gets the identities}
\begin{equation}
\begin{cases}
	\displaystyle \bm{G} + \frac{m_{ij}}{m_i}\bm{g}
	= \bm{G}^\prime + \frac{\epsilon_j}{\epsilon_i + \epsilon_j}\bm{g},\\
	\displaystyle \bm{G} - \frac{m_{ij}}{m_j}\bm{g}
	= \bm{G}^\prime - \frac{\epsilon_i}{\epsilon_i + \epsilon_j}\bm{g}.
\end{cases}
\end{equation}
Thus, we can rewrite Eq.\ \eqref{eq:fi_fj} as
\begin{align}
	f_i\left(\bm{V}_1\right) f_j\left(\bm{V}_2\right)
	&= n_i n_j  v_{\rm T}^{-3} \left(\epsilon_i \epsilon_j\right)^{3/2}
	\pi^{-3} \exp\left[-(\epsilon_i + \epsilon_j)G^{\prime2} -\frac{\epsilon_i \epsilon_j}{\epsilon_i + \epsilon_j}g^2\right]\nonumber\\
	&\hspace{1em}\times
	\left[1+\epsilon_i \Pi^{(i)}_{\alpha\beta}
		\left(G^\prime_\alpha+\frac{\epsilon_j}{\epsilon_i + \epsilon_j}g_\alpha\right)
		\left(G^\prime_\beta+\frac{\epsilon_j}{\epsilon_i + \epsilon_j}g_\beta\right)\right.\nonumber\\
	&\hspace{2.5em}\left.	
		+\epsilon_j \Pi^{(j)}_{\alpha\beta}
		\left(G^\prime_\alpha-\frac{\epsilon_i}{\epsilon_i + \epsilon_j}g_\alpha\right)
		\left(G^\prime_\beta-\frac{\epsilon_i}{\epsilon_i + \epsilon_j}g_\beta\right)\right].
	\label{eq:fi_fj_linear}
\end{align}

Let us rewrite Eq.\ \eqref{eq:Lambda_ij}.
From Eqs.\ \eqref{eq:inelastic_collision} and \eqref{eq:G_prime}, one gets
\begin{align}
	&m_i v_{1,\alpha}^\prime v_{1,\beta}^\prime - m_i v_{1,\alpha} v_{1,\beta}\nonumber\\
	&= -m_{ij}(1+e_{ij})v_{\rm T}^2 (\bm{g}\cdot \widehat{\bm{\sigma}})
		\left[\left(G_\alpha+\frac{m_{ij}}{m_i}g_\alpha\right)\widehat{\sigma}_\beta
			+ \left(G_\beta+\frac{m_{ij}}{m_i}g_\beta\right)\widehat{\sigma}_\alpha
			-\frac{m_{ij}}{m_i}(1+e_{ij})(\bm{g}\cdot \widehat{\bm{\sigma}})\widehat{\sigma}_\alpha \widehat{\sigma}_\beta\right]\nonumber\\
	&= -m_{ij}(1+e_{ij})v_{\rm T}^2 (\bm{g}\cdot \widehat{\bm{\sigma}})
		\left[G^\prime_\alpha \widehat{\sigma}_\beta + G^\prime_\beta \widehat{\sigma}_\alpha
			+\frac{\epsilon_j}{\epsilon_i+\epsilon_j}(g_\alpha \widehat{\sigma}_\beta + g_\beta \widehat{\sigma}_\alpha)
			-\frac{m_{ij}}{m_i}(1+e_{ij})(\bm{g}\cdot \widehat{\bm{\sigma}})\widehat{\sigma}_\alpha \widehat{\sigma}_\beta\right].
	\label{eq:mv1pv1p-mv1v1}
\end{align}
Using Eqs.\ \eqref{eq:fi_fj} and \eqref{eq:mv1pv1p-mv1v1}, we can rewrite the collisional moment $\Lambda^{(ij)}_{\alpha\beta}$ as
\begin{align}
	\Lambda^{(ij)}_{\alpha\beta}
	&= m_{ij} (1+e_{ij})n_i n_j \sigma^{(ij)2}(\epsilon_i\epsilon_j)^{3/2} v_{\rm T}^3 \widetilde{\Lambda}^{(ij)}_{\alpha\beta},
	\label{eq:Lambda_dimensionless}
\end{align}
with the linear collisional moment
\begin{align}
	\widetilde{\Lambda}^{(ij)}_{\alpha\beta}
	&\equiv \pi^{-3}
	\int d\bm{G}^\prime \int d\bm{g} \int d\widehat{\bm{\sigma}}
	\Theta(\widehat{\bm{\sigma}}\cdot \bm{g}) (\widehat{\bm{\sigma}}\cdot \bm{g})^2
	\exp\left[-(\epsilon_i + \epsilon_j)G^{\prime2} -\frac{\epsilon_i \epsilon_j}{\epsilon_i + \epsilon_j}g^2\right]\nonumber\\
	&\hspace{1em}\times
	\left[G^\prime_\alpha \widehat{\sigma}_\beta + G^\prime_\beta \widehat{\sigma}_\alpha
			+\frac{\epsilon_j}{\epsilon_i+\epsilon_j}(g_\alpha \widehat{\sigma}_\beta + g_\beta \widehat{\sigma}_\alpha)
			-\frac{m_{ij}}{m_i}(1+e_{ij})(\bm{g}\cdot \widehat{\bm{\sigma}})\widehat{\sigma}_\alpha \widehat{\sigma}_\beta\right]\nonumber\\
	&\hspace{1em}\times
	\left[1+\epsilon_i \Pi^{(i)}_{\gamma\delta}
		\left(G^\prime_\gamma+\frac{\epsilon_j}{\epsilon_i + \epsilon_j}g_\gamma\right)
		\left(G^\prime_\delta+\frac{\epsilon_j}{\epsilon_i + \epsilon_j}g_\delta\right)\right.\nonumber\\
	&\hspace{2.5em}\left.	
		+\epsilon_j \Pi^{(j)}_{\gamma\delta}
		\left(G^\prime_\gamma-\frac{\epsilon_i}{\epsilon_i + \epsilon_j}g_\gamma\right)
		\left(G^\prime_\delta-\frac{\epsilon_i}{\epsilon_i + \epsilon_j}g_\delta\right)\right].
	\label{eq:tilde_Lambda}
\end{align}
For further calculation, let us first introduce $\widetilde{I}_{ij}^{(\ell)}(\widehat{\bm{\sigma}})$ and $\widetilde{I}_{ij,\alpha}^{(\ell)}(\widehat{\bm{\sigma}})$ as
\begin{align}
	\begin{Bmatrix} \widetilde{I}_{ij}^{(\ell)}(\widehat{\bm{\sigma}}) \\ \widetilde{I}_{ij,\alpha}^{(\ell)}(\widehat{\bm{\sigma}}) \end{Bmatrix}
	&\equiv \frac{1}{\pi^3} \int d\bm{G}^\prime \int d\bm{g}
		\Theta(\widehat{\bm{\sigma}}\cdot \bm{g}) (\widehat{\bm{\sigma}}\cdot \bm{g})^\ell
		\begin{Bmatrix} 1 \\ g_\alpha \end{Bmatrix}
		\exp\left[-(\epsilon_i + \epsilon_j)G^{\prime2} -\frac{\epsilon_i \epsilon_j}{\epsilon_i + \epsilon_j}g^2\right]\nonumber\\
	&\hspace{1em}\times
	\left[1+\epsilon_i \Pi^{(i)}_{\gamma\delta}
		\left(G^\prime_\gamma+\frac{\epsilon_j}{\epsilon_i + \epsilon_j}g_\gamma\right)
		\left(G^\prime_\delta+\frac{\epsilon_j}{\epsilon_i + \epsilon_j}g_\delta\right)\right.\nonumber\\
	&\hspace{2.5em}\left.	
		+\epsilon_j \Pi^{(j)}_{\gamma\delta}
		\left(G^\prime_\gamma-\frac{\epsilon_i}{\epsilon_i + \epsilon_j}g_\gamma\right)
		\left(G^\prime_\delta-\frac{\epsilon_i}{\epsilon_i + \epsilon_j}g_\delta\right)\right]\nonumber\\
	&= \frac{1}{\pi^{3/2}(\epsilon_i + \epsilon_j)^{3/2}} \int d\bm{g}
		\Theta(\widehat{\bm{\sigma}}\cdot \bm{g}) (\widehat{\bm{\sigma}}\cdot \bm{g})^\ell
		\begin{Bmatrix} 1 \\ g_\alpha \end{Bmatrix}
		\exp\left(-\frac{\epsilon_i \epsilon_j}{\epsilon_i + \epsilon_j}g^2\right)\mathcal{P}_1(\{g\}),
	\label{eq:tilde_I}
\end{align}
with
\begin{align}
	\mathcal{P}_1(\{g\})
	&\equiv 1
		+ \frac{\epsilon_i \epsilon_j^2}{(\epsilon_i+\epsilon_j)^2}g_\gamma g_\delta \Pi^{(i)}_{\gamma\delta}
		+ \frac{\epsilon_i^2 \epsilon_j}{(\epsilon_i+\epsilon_j)^2}g_\gamma g_\delta \Pi^{(j)}_{\gamma\delta}.
\end{align}
We also introduce $\widehat{I}_{ij,\alpha}^{(\ell)}(\widehat{\bm{\sigma}})$ as
\begin{align}
	\widehat{I}_{ij,\alpha}^{(\ell)}(\widehat{\bm{\sigma}})
	&\equiv \frac{1}{\pi^3} \int d\bm{G}^\prime \int d\bm{g}
		\Theta(\widehat{\bm{\sigma}}\cdot \bm{g}) (\widehat{\bm{\sigma}}\cdot \bm{g})^\ell
		G^\prime_\alpha
		\exp\left[-(\epsilon_i + \epsilon_j)G^{\prime2} -\frac{\epsilon_i \epsilon_j}{\epsilon_i + \epsilon_j}g^2\right]\nonumber\\
	&\hspace{1em}\times
	\left[1+\epsilon_i \Pi^{(i)}_{\gamma\delta}
		\left(G^\prime_\gamma+\frac{\epsilon_j}{\epsilon_i + \epsilon_j}g_\gamma\right)
		\left(G^\prime_\delta+\frac{\epsilon_j}{\epsilon_i + \epsilon_j}g_\delta\right)\right.\nonumber\\
	&\hspace{2.5em}\left.	
		+\epsilon_j \Pi^{(j)}_{\gamma\delta}
		\left(G^\prime_\gamma-\frac{\epsilon_i}{\epsilon_i + \epsilon_j}g_\gamma\right)
		\left(G^\prime_\delta-\frac{\epsilon_i}{\epsilon_i + \epsilon_j}g_\delta\right)\right]\nonumber\\
	&= \frac{1}{\pi^{3/2}(\epsilon_i + \epsilon_j)^{3/2}} \int d\bm{g}
		\Theta(\widehat{\bm{\sigma}}\cdot \bm{g}) (\widehat{\bm{\sigma}}\cdot \bm{g})^\ell
		\exp\left(-\frac{\epsilon_i \epsilon_j}{\epsilon_i + \epsilon_j}g^2\right)\mathcal{Q}_{1,\alpha}(\{g\}),
	\label{eq:hat_I}
\end{align}
with
\begin{align}
	\mathcal{Q}_{1,\alpha}(\{g\})
	&= \frac{\epsilon_i \epsilon_j}{(\epsilon_i + \epsilon_j)^2} g_\gamma
		\left(\Pi^{(i)}_{\alpha\gamma} - \Pi^{(j)}_{\alpha\gamma}\right).
\end{align}
Using Eqs.\ \eqref{eq:tilde_I} and \eqref{eq:hat_I}, Eq.\ \eqref{eq:tilde_Lambda} is rewritten as
\begin{align}
	\widetilde{\Lambda}^{(ij)}_{\alpha\beta}
	&\equiv
	\int d\widehat{\bm{\sigma}}
	\left\{\widehat{I}^{(2)}_{ij, \alpha}(\widehat{\bm{\sigma}}) \widehat{\sigma}_\beta
		+ \widehat{I}^{(2)}_{ij, \beta}(\widehat{\bm{\sigma}}) \widehat{\sigma}_\alpha\right.\nonumber\\
	&\hspace{4.5em}\left.
		+ \frac{\epsilon_j}{\epsilon_i+\epsilon_j}\left[
			\widetilde{I}^{(2)}_{ij, \alpha}(\widehat{\bm{\sigma}}) \widehat{\sigma}_\beta
			+ \widetilde{I}^{(2)}_{ij, \beta}(\widehat{\bm{\sigma}}) \widehat{\sigma}_\alpha
		\right]
		- \frac{m_{ij}}{m_i}(1+e_{ij})\widetilde{I}^{(3)}_{ij}(\widehat{\bm{\sigma}}) \widehat{\sigma}_\alpha \widehat{\sigma}_\beta
	\right\}.
	\label{eq:tilde_Lambda_2}
\end{align}
Let us go further by integrating over $\widehat{\bm{\sigma}}$.
Here, the {following results are needed}:
\begin{align}
\label{sigma}
	\int d\widehat{\bm{\sigma}}
	\Theta(\widehat{\bm{\sigma}}\cdot \bm{g}) (\widehat{\bm{\sigma}}\cdot \bm{g})^n
	\widehat{\bm{\sigma}}
	&= \beta_{n+1} g^{n-1} \bm{g},\\
\label{sigma1}
	\int d\widehat{\bm{\sigma}}
	\Theta(\widehat{\bm{\sigma}}\cdot \bm{g}) (\widehat{\bm{\sigma}}\cdot \bm{g})^n
	\widehat{\sigma}_\alpha \widehat{\sigma}_\beta
	&= \frac{\beta_n}{n+3}g^{n-2} (ng_\alpha g_\beta + g^2 \delta_{\alpha\beta}),
\end{align}
with
\begin{equation}
	\beta_n = \pi \frac{\displaystyle \Gamma\left(\frac{n+1}{2}\right)}{\displaystyle \Gamma\left(\frac{n+3}{2}\right)}
	=\frac{2\pi}{n+1}.
	\label{eq:def_beta_n}
\end{equation}
With the aid of {Eqs.\ \eqref{sigma}--\eqref{sigma1}}, one gets
\begin{align}
	\int d\widehat{\bm{\sigma}}
	\widehat{I}^{(2)}_{ij, \alpha}(\widehat{\bm{\sigma}}) \widehat{\sigma}_\beta
	&= \frac{\beta_3}{\pi^{3/2}(\epsilon_i + \epsilon_j)^{3/2}} \int d\bm{g}
		g g_\beta \exp\left(-\frac{\epsilon_i \epsilon_j}{\epsilon_i + \epsilon_j}g^2\right)\mathcal{Q}_{1,\alpha}(\{g\})\nonumber\\
	&= \frac{2\sqrt{\pi}}{3} (\epsilon_i \epsilon_j)^{-3/2}
	\frac{1}{\sqrt{\epsilon_i \epsilon_j (\epsilon_i + \epsilon_j)}}
	\left(\Pi^{(i)}_{\alpha\beta} - \Pi^{(j)}_{\alpha\beta}\right),
\end{align}
and there,
\begin{equation}
	\int d\widehat{\bm{\sigma}}
	\widehat{I}^{(2)}_{ij, \alpha}(\widehat{\bm{\sigma}}) \widehat{\sigma}_\beta
	+ \int d\widehat{\bm{\sigma}}
	\widehat{I}^{(2)}_{ij, \beta}(\widehat{\bm{\sigma}}) \widehat{\sigma}_\alpha
	= \frac{2\sqrt{\pi}}{3} (\epsilon_i \epsilon_j)^{-3/2}
		\left(\frac{\epsilon_i + \epsilon_j}{\epsilon_i \epsilon_j}\right)^{3/2}
		\frac{2\epsilon_i\epsilon_j}{(\epsilon_i + \epsilon_j)^2}
		\left(\Pi^{(i)}_{\alpha\beta} - \Pi^{(j)}_{\alpha\beta}\right).
	\label{eq:linear_1}
\end{equation}

Similarly, one achieves the result
\begin{align}
	\int d\widehat{\bm{\sigma}}
	\widetilde{I}^{(2)}_{ij, \alpha}(\widehat{\bm{\sigma}}) \widehat{\sigma}_\beta
	&= \frac{\beta_3}{\pi^{3/2}(\epsilon_i + \epsilon_j)^{3/2}} \int d\bm{g}
		g g_\alpha g_\beta \exp\left(-\frac{\epsilon_i \epsilon_j}{\epsilon_i + \epsilon_j}g^2\right)\mathcal{P}_1(\{g\})\nonumber\\
	&= \frac{2\sqrt{\pi}}{3} (\epsilon_i \epsilon_j)^{-3/2}
		\left(\frac{\epsilon_i + \epsilon_j}{\epsilon_i \epsilon_j}\right)^{3/2}
		\left[\delta_{\alpha\beta}
			+ \frac{6}{5}\left(\frac{\epsilon_j}{\epsilon_i+\epsilon_j}\Pi^{(i)}_{\alpha\beta}
				+ \frac{\epsilon_i}{\epsilon_i+\epsilon_j}\Pi^{(j)}_{\alpha\beta}\right)\right],
\end{align}
and then,
\begin{align}
	&\int d\widehat{\bm{\sigma}}
	\frac{\epsilon_j}{\epsilon_i+\epsilon_j}\left[
		\widetilde{I}^{(2)}_{ij, \alpha}(\widehat{\bm{\sigma}}) \widehat{\sigma}_\beta
		+ \widetilde{I}^{(2)}_{ij, \beta}(\widehat{\bm{\sigma}}) \widehat{\sigma}_\alpha
	\right]\nonumber\\
	&= \frac{2\sqrt{\pi}}{3} (\epsilon_i \epsilon_j)^{-3/2}
		\left(\frac{\epsilon_i + \epsilon_j}{\epsilon_i \epsilon_j}\right)^{3/2}
		\frac{2\epsilon_j}{\epsilon_i+\epsilon_j}
		\left[\delta_{\alpha\beta}
			+ \frac{6}{5}\left(\frac{\epsilon_j}{\epsilon_i+\epsilon_j}\Pi^{(i)}_{\alpha\beta}
				+ \frac{\epsilon_i}{\epsilon_i+\epsilon_j}\Pi^{(j)}_{\alpha\beta}\right)\right].\label{eq:linear_2}
\end{align}

In addition, one gets
\begin{align}
	\int d\widehat{\bm{\sigma}}
	\widetilde{I}^{(3)}_{ij}(\widehat{\bm{\sigma}}) \widehat{\sigma}_\alpha \widehat{\sigma}_\beta
	&= \frac{\beta_3}{6\pi^{3/2}(\epsilon_i + \epsilon_j)^{3/2}} \int d\bm{g}
		(g^3 \delta_{\alpha\beta} + 3g g_\alpha g_\beta) \exp\left(-\frac{\epsilon_i \epsilon_j}{\epsilon_i + \epsilon_j}g^2\right)\mathcal{P}_1(\{g\})\nonumber\\
	&= \frac{2\sqrt{\pi}}{3} (\epsilon_i \epsilon_j)^{-3/2}
		\left(\frac{\epsilon_i + \epsilon_j}{\epsilon_i \epsilon_j}\right)^{3/2}
		\left[\delta_{\alpha\beta}
			+ \frac{3}{5}\left(\frac{\epsilon_j}{\epsilon_i+\epsilon_j}\Pi^{(i)}_{\alpha\beta}
				+ \frac{\epsilon_i}{\epsilon_i+\epsilon_j}\Pi^{(j)}_{\alpha\beta}\right)\right],
\end{align}
and then
\begin{align}
	&\int d\widehat{\bm{\sigma}}
	\frac{m_{ij}}{m_i}(1+e_{ij})\widetilde{I}^{(3)}_{ij}(\widehat{\bm{\sigma}}) \widehat{\sigma}_\alpha \widehat{\sigma}_\beta\nonumber\\
	&= \frac{2\sqrt{\pi}}{3} (\epsilon_i \epsilon_j)^{-3/2}
		\left(\frac{\epsilon_i + \epsilon_j}{\epsilon_i \epsilon_j}\right)^{3/2}
		\frac{m_{ij}}{m_i}(1+e_{ij})
		\left[\delta_{\alpha\beta}
			+ \frac{3}{5}\left(\frac{\epsilon_j}{\epsilon_i+\epsilon_j}\Pi^{(i)}_{\alpha\beta}
				+ \frac{\epsilon_i}{\epsilon_i+\epsilon_j}\Pi^{(j)}_{\alpha\beta}\right)\right].
	\label{eq:linear_3}
\end{align}

Substituting Eqs.\ \eqref{eq:linear_1}, \eqref{eq:linear_2}, and \eqref{eq:linear_3} into Eq.\ \eqref{eq:tilde_Lambda_2}, one obtains
\begin{align}
	\widetilde{\Lambda}^{(ij)}_{\alpha\beta}
	&=\frac{2\sqrt{\pi}}{3} (\epsilon_i \epsilon_j)^{-3/2}
		\left(\frac{\epsilon_i + \epsilon_j}{\epsilon_i \epsilon_j}\right)^{3/2}\nonumber\\
	&\hspace{1em}\times\left(\left[\frac{2\epsilon_j}{\epsilon_i+\epsilon_j}-\frac{m_{ij}}{m_i}(1+e_{ij})\right]\delta_{\alpha\beta}
		 +\frac{2\epsilon_i\epsilon_j}{(\epsilon_i+\epsilon_j)^2}\left\{1+\frac{3}{5}\frac{\epsilon_i+\epsilon_j}{\epsilon_i}\left[\frac{2\epsilon_j}{\epsilon_i+\epsilon_j}
		-\frac{1}{2}\frac{m_{ij}}{m_i}(1+e_{ij})\right]\Pi^{(i)}_{\alpha\beta}\right\}\right.\nonumber\\
	&\hspace{2.5em}\left.-\frac{2\epsilon_i\epsilon_j}{(\epsilon_i+\epsilon_j)^2}
		\left\{1-\frac{3}{5}\frac{\epsilon_i+\epsilon_j}{\epsilon_j}\left[\frac{2\epsilon_j}{\epsilon_i+\epsilon_j}
		-\frac{1}{2}\frac{m_{ij}}{m_i}(1+e_{ij})\right]\Pi^{(j)}_{\alpha\beta}\right\}\right)\nonumber\\
	&= \frac{2\sqrt{\pi}}{3} (\epsilon_i \epsilon_j)^{-3/2}
		\left(\frac{\epsilon_i + \epsilon_j}{\epsilon_i \epsilon_j}\right)^{3/2}
		\left\{\left[\lambda_{ij}-\frac{1}{2}
		\frac{m_{ij}}{m_i}(1+e_{ij})\right]\delta_{\alpha\beta} \right.\nonumber\\
	&\hspace{1em}\left. +2\frac{\epsilon_i \epsilon_j}{(\epsilon_i+\epsilon_j)^2}
		\left[\left(1+\frac{3}{5}\frac{\epsilon_i+\epsilon_j}{\epsilon_i}\lambda_{ij}\right)\Pi^{(i)}_{\alpha\beta}
		-\left(1-\frac{3}{5}\frac{\epsilon_i+\epsilon_j}{\epsilon_j}\lambda_{ij}\right)\Pi^{(j)}_{\alpha\beta}\right]\right\}.
	\label{eq:linear_sum}
\end{align}

Finally, the combination of Eqs.\ \eqref{eq:Lambda_dimensionless} and \eqref{eq:linear_sum} yields Eq.\ \eqref{eq:Lambda_ij_alphabeta}.

\section{Detailed flow curves}\label{sec:detailed_rheology}
In this Appendix, we present supplemental results of rheology explained in Sec.\ \ref{sec:comparison} of the main text.
We  {display} the results for $\theta$ versus {$\dot\gamma^*$} and $\eta^*\equiv-(\nu_1\Pi_{xy}^{(1)*}+\nu_2\Pi_{xy}^{(2)*})/\dot\gamma^*$ versus {$\dot\gamma^*$}.

When we focus on the reduced temperature $\theta$ (see Fig.\ \ref{fig:theta_shear}), the effect of the bidispersity only appears around an intermediate shear regime ($\dot\gamma^*\simeq 5.0$), where the discontinuous change corresponding to the DST is observed.
Although this discontinuous change itself is reported even in monodisperse systems \cite{DST16}, the point at which the discontinuous change occurs depends on the size ratio.
It is noteworthy that the change of the reduced temperature is drastic but continuous when the size ratio becomes large (see the data for $\sigma^{(1)}/\sigma^{(2)}=2.0$ and $5.0$ in Fig.\ \ref{fig:theta_shear}).

As well as in Fig.\ \ref{fig:theta_shear}, the viscosity $\eta^*$ is also plotted against the shear rate $\dot\gamma^*$ in Fig.\ \ref{fig:eta_shear}.
If the size ratio $\sigma^{(1)}/\sigma^{(2)}$ is close to unity, such as $1.4$, the flow curves of $\theta$ and $\eta^*$ are similar to the corresponding ones for monodisperse gases, in which there are discontinuous changes of $\theta$ and $\eta^*$ around $\gamma^*\approx 5$.
However, as the size ratio increases, the discontinuous changes of $\theta$ and $\eta^*$ become continuous.
Moreover, these flow curves for inelastic inertial suspensions for large $\sigma^{(1)}/\sigma^{(2)}$ are characteristic.
Indeed, the slopes of $\theta$ and $\eta^*$ are oscillated with $\dot\gamma^*$ before reaching their asymptotic values in the large shear rate limit.

\begin{figure}[htbp]
	\centering
	\includegraphics[width=\linewidth]{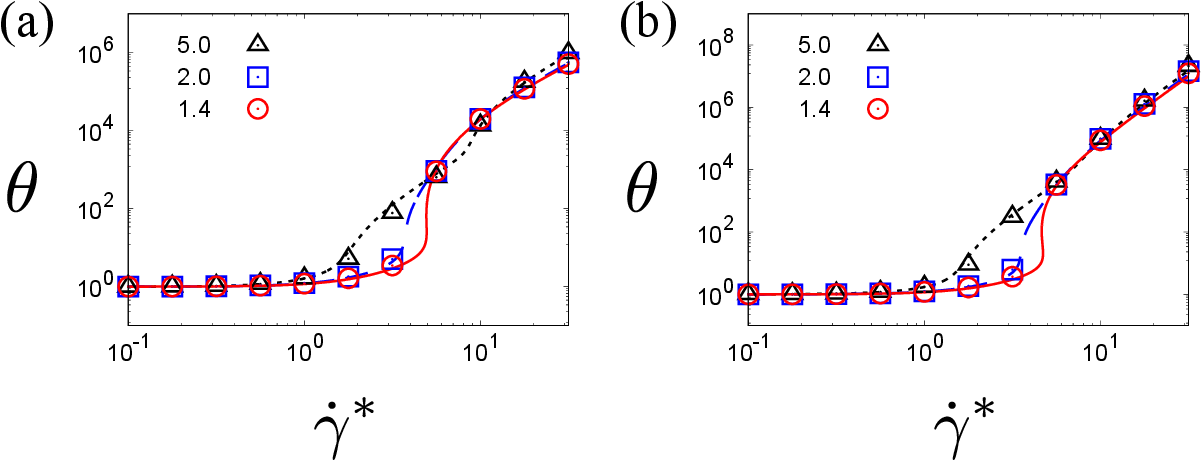}
	\caption{
	(a) Dimensionless temperature $\theta$ against the dimensionless shear rate
	$\dot\gamma^*$ for $\sigma^{(1)}/\sigma^{(2)}=1.4$ (solid line and open circles), $2.0$ (dashed line and open squares), and $5.0$ (dotted line and open triangles), when we fix $\varphi=0.01$, $\xi_{\rm env}=1.0$, and $\nu_1=\nu_2=1/2$ for $e=0.9$.
 (b) $\theta$ against $\dot\gamma^*$ for $\sigma^{(1)}/\sigma^{(2)}=1.4$, 2.0 and 5.0 by fixing $\varphi=0.01$, $\xi_{\rm env}=1.0$, and $\nu_1=\nu_2=1/2$ for $e=1.0$.
	The lines and symbols correspond to the steady solutions of the theoretical predictions \eqref{eq:theta_vartheta_gammadot} and the simulation results, respectively.}
	\label{fig:theta_shear}
\end{figure}

\begin{figure}[htbp]
	\centering
	\includegraphics[width=\linewidth]{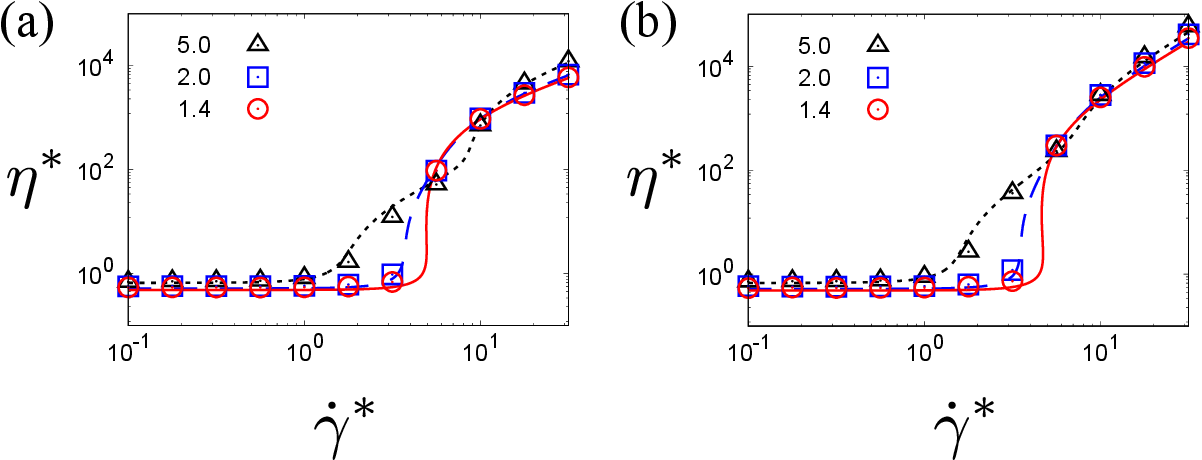}
	\caption{The dimensionless viscosity $\eta^*$ against the dimensionless shear rate $\dot\gamma^*$ for $\sigma^{(1)}/\sigma^{(2)}=1.4$ (solid line and open circles), $2.0$ (dashed line and open squares), and $5.0$ (dotted line and open triangles) when we fix $\varphi=0.01$, $\xi_{\rm env}=1.0$, and $\nu_1=\nu_2=1/2$ for (a) $e=0.9$ and (b) $1$.
	The lines and symbols correspond to the steady solutions of the theoretical predictions \eqref{eq:eta_i_gammadot} and the simulation results, respectively.}
	\label{fig:eta_shear}
\end{figure}

We also draw 3D-phase diagrams of the number of solutions obtained by the kinetic theory in the ($\nu_1, \dot\gamma^*, e$)-plane for $\varphi=0.01$ in Fig.~\ref{fig:phase_diagram}.
The filled regions represent those whose number of solutions is three, while the empty regions represent only one solution.
These plots show that the regions for the multiple solutions are localized in the narrow regimes in the ($\nu_1, \dot\gamma^*, e$)-plane.

\begin{figure}[htbp]
	\centering
	\includegraphics[width=\linewidth]{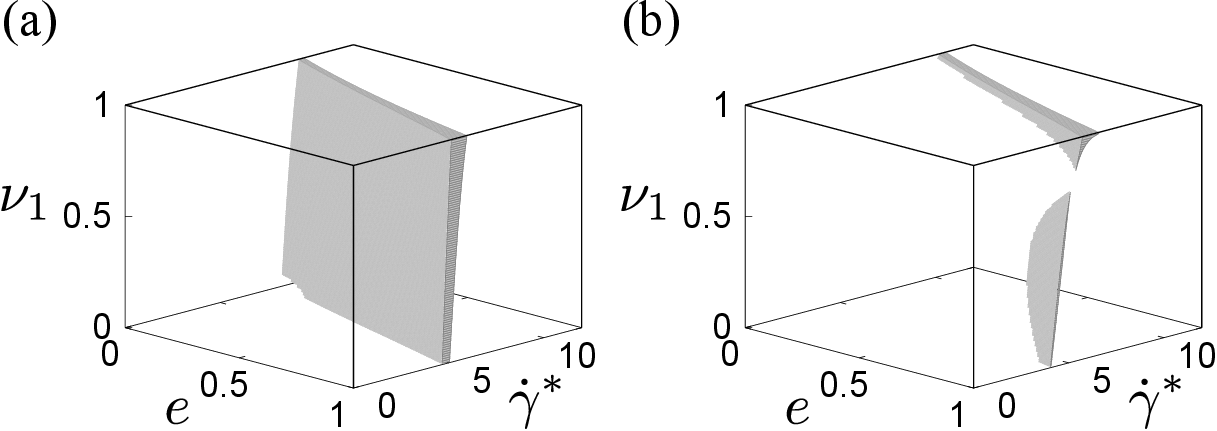}
	\caption{
	Phase diagrams of the number of solutions against ($e$, $\nu_1$, $\dot\gamma^*$) for (a) $\sigma^{(1)}/\sigma^{(2)}=1.1$ and (b) $1.4$ by fixing $\varphi=0.01$ and $\xi_{\rm env}=1.0$.
	Here, the filled (empty) region represents that the number of the solutions is three (unity).}
	\label{fig:phase_diagram}
\end{figure}

\section{Appearance/disappearance of the discontinuous transition}\label{sec:appear_disappear}
In this Appendix, let us show how the discontinuous transition appears/disappears when we change the parameters of the mixture.
This appendix consists of three subsections.
In the first part, we discuss how the results depend on the environmental temperature $\xi_\mathrm{env}$.
In the second part, we distinguish the region of DST-like behavior from the CST-like behavior when we fix $\nu_1=\nu_2=1/2$.
In the last part, we also distinguish the region of DST-like behavior from the CST-like behavior if we fix the volume ratio $\mathcal{V}=1$,

\subsection{Effect of the environmental temperature $\xi_\mathrm{env}$}
First, since $\xi_\mathrm{env}\propto \sqrt{T_\mathrm{env}}$, we analyze the dependence of the flow curves on the environmental temperature for $\nu_1=\nu_2=1/2$. 
This temperature determines the state in the low shear regime, but is independent in the high shear regime.
The latter fact is understood because interparticle collisions are dominant in the latter regime.
Figure \ref{fig:xi_change} illustrates the above fact:
The high shear regime is independent of the choice of the environmental temperature, but the low shear regime is determined by the value of the environmental temperature.
It is interesting to note that the Newtonian regime becomes narrower as $\xi_\mathrm{env}$ increases. 
More importantly, DST-like behavior for $\eta^*$ for low $\xi_\mathrm{env}$ becomes CST-like as $\xi_\mathrm{env}$ increases.
\begin{figure}[htbp]
	\centering
	\includegraphics[width=0.5\linewidth]{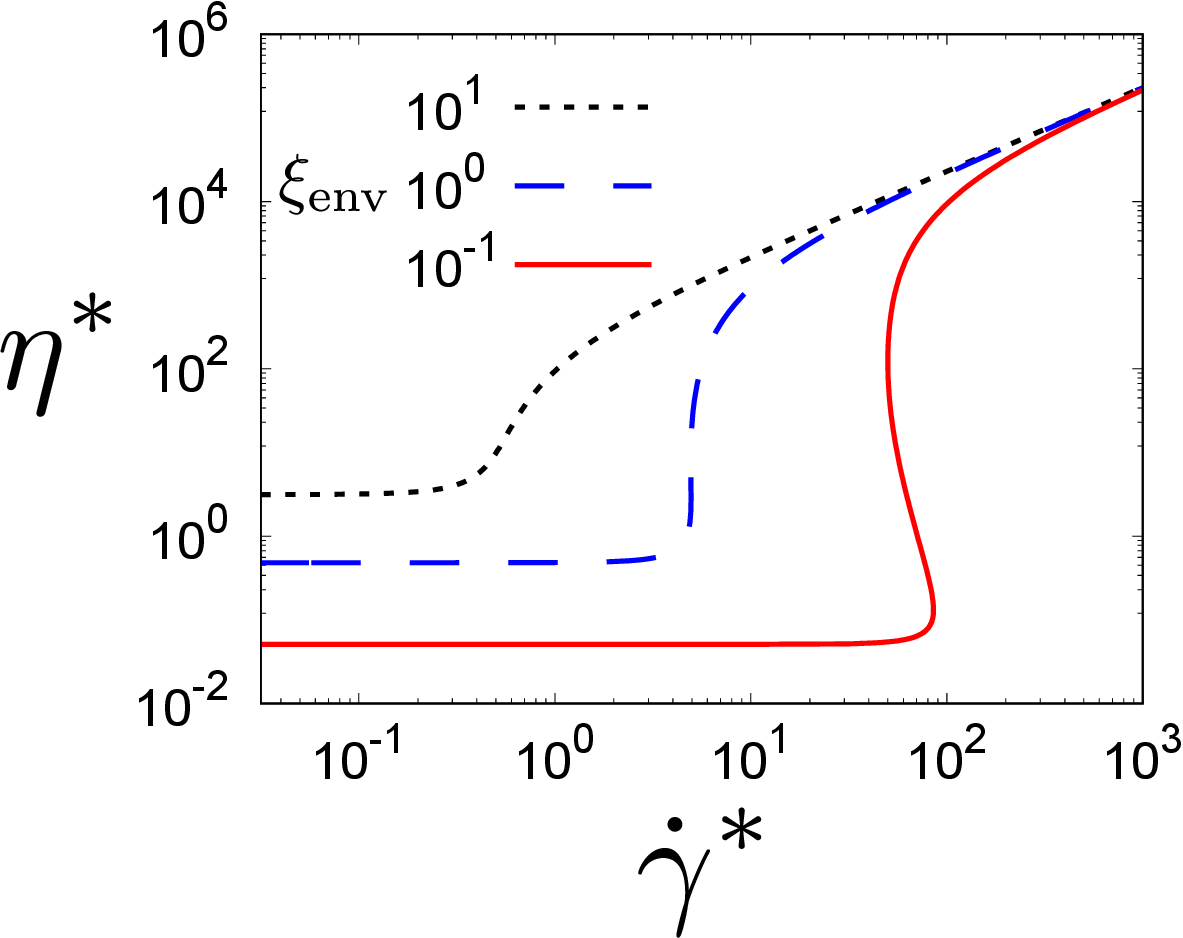}
	\caption{Plots of the global viscosity $\eta^*$ against the dimensionless shear rate $\dot\gamma^*$ for $\xi_{\rm env}=10^{-1}$ (solid line), $10^0$ (dashed line), and $10^1$ (dotted line) when we fix $\varphi=0.01$, $e=0.9$, $\sigma^{(1)}/\sigma^{(2)}=1.4$, and $\nu_1=\nu_2=1/2$.}
	\label{fig:xi_change}
\end{figure}

\subsection{Effect of the size ratio for $N_1=N_2$}\label{sec:size_ratio_N1_N2}
\begin{figure}[htbp]
	\centering
	\includegraphics[width=0.5\linewidth]{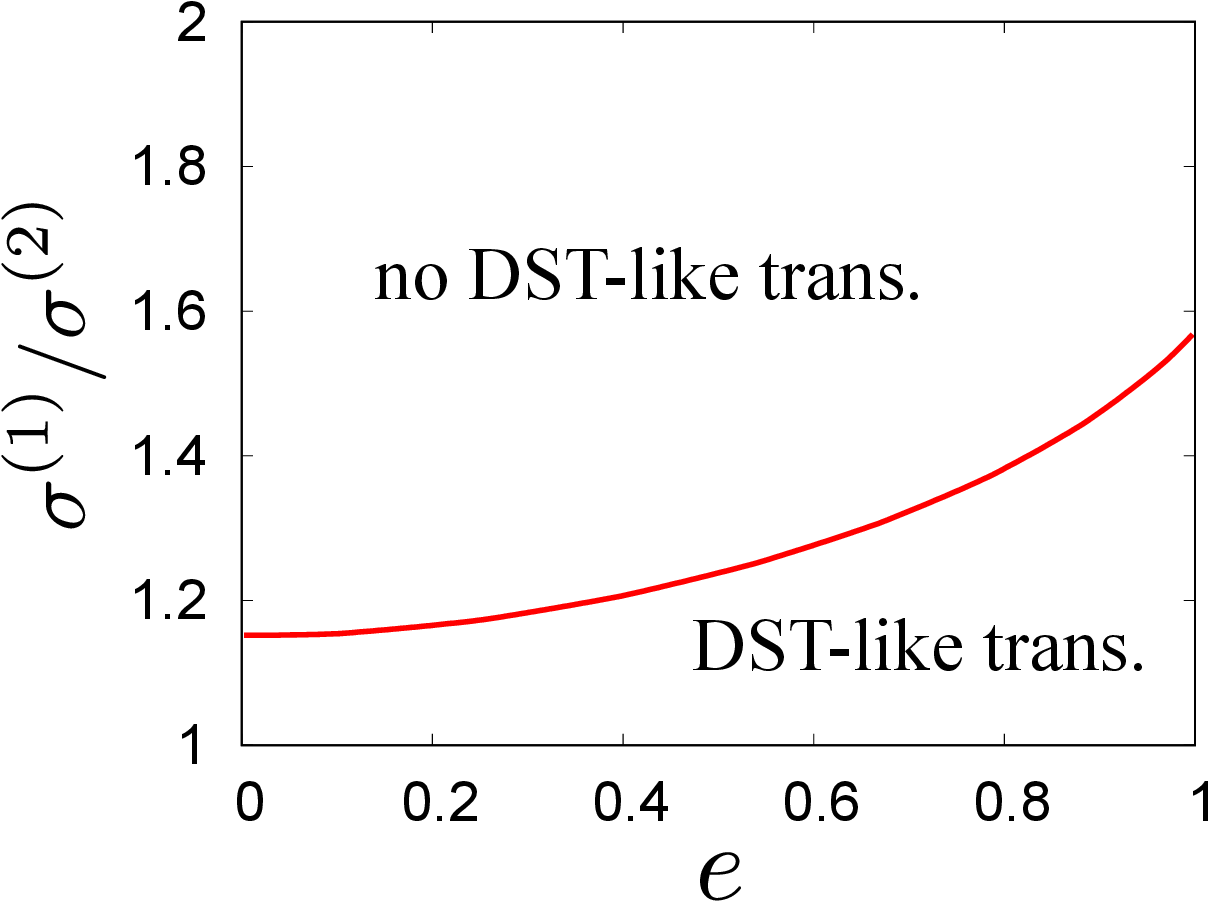}
	\caption{Plot of the critical size ratio against the restitution coefficient when we fix $\varphi=0.01$, $\xi_{\rm env}=1.0$, and $\nu_1=\nu_2=1/2$.}
	\label{fig:critical_sigma_ratio_N}
\end{figure}
Next, let us consider the size ratio dependence in the case of $\nu_1=\nu_2=1/2$ based on the theoretical calculation.
In this case, the discontinuous jumps are observed when the size ratio is not large such as $\sigma^{(1)}/\sigma^{(2)}=1.4$ in Figs.~\ref{fig:vartheta_shear} and \ref{fig:eta_ratio_shear} as shown in Fig.~\ref{fig:critical_sigma_ratio_N}. 
On the other hand, the flow curves become continuous for larger size ratio. 
We can understand this behavior by considering first the discontinuous jump for the monodisperse system ($\nu_1=1$, $\nu_2=0$).
Depending on the value of the (reduced) shear rate $\dot\gamma^*\equiv \dot\gamma / \zeta_1$, there are two different regimes; high-shear and low-shear regimes.
The former regime is known as Bagnold's expression, $\eta^*\propto \dot\gamma^*/(\xi_\mathrm{env}^2\varphi^2)$ for $e<1$ \cite{DST16}.
We note that, for the elastic case, a different expression is obtained as $\eta^*\propto \dot\gamma^{*2}$.
However, the latter regime (low shear regime) is determined by the interaction between the particles and the solvent \cite{DST16}, and so $\eta^*\sim 1$.
These two regimes switch to each other at $\dot\gamma^*\simeq 1$.
Given that the difference between two regimes is proportional to the inverse of the volume fraction, the flow curve forms an S-shape connecting the two regimes.

Now, we consider binary systems.
If the size ratio is not sufficiently large, such as $\sigma^{(1)}/\sigma^{(2)}=1.4$ as shown in Figs.~\ref{fig:theta_shear} and \ref{fig:eta_shear}, the picture for the monodisperse system can also be used for a binary system.
This means that the discontinuous jumps appear in this case.
On the other hand, as the size ratio increases, collisions between smaller and larger particles compete with those between particles with the same size.
This means that we need to discuss the mixing energy between smaller and larger particles in this case.
Relating to this, we may use a discussion analogous to the phase coexistence and spinodal lines at equilibrium phase transitions, respectively, in the phase space of ($\theta$, $\dot\gamma^*$, $\sigma^{(1)}/\sigma^{(2)}$).
\begin{figure}[htbp]
	\centering
	\includegraphics[width=0.5\linewidth]{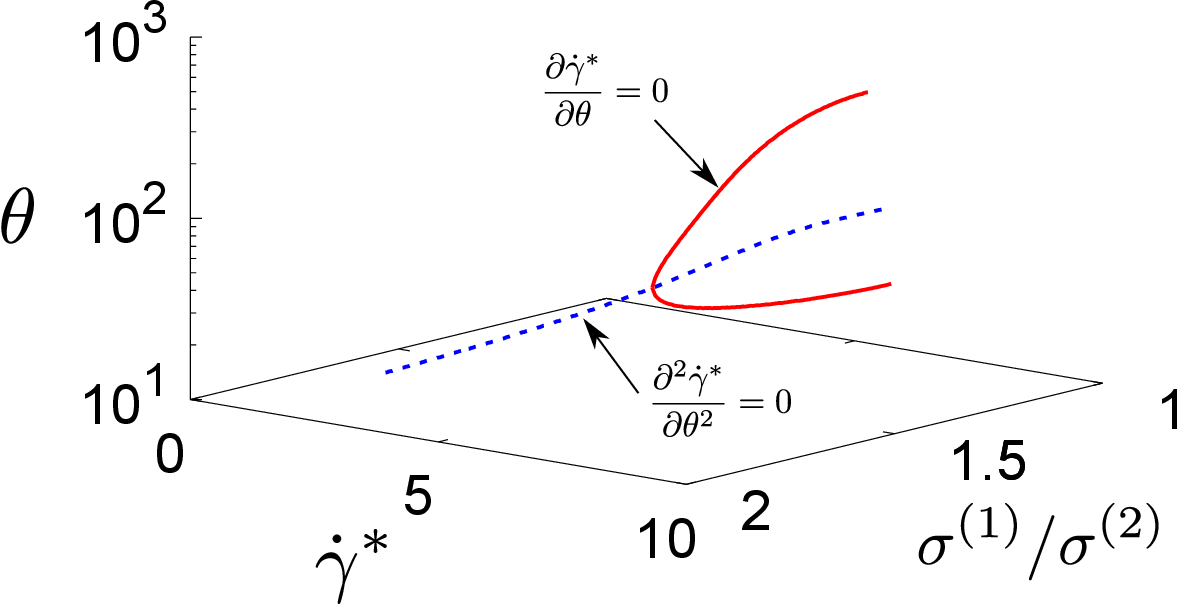}
	\caption{
	Plot of the phase coexistence line $\partial \dot\gamma^*/\partial \theta=0$ (solid lines) and the spinodal line $\partial^2 \dot\gamma^*/\partial \theta^2=0$ (dashed line) for $\varphi=0.01$, $e=0.9$ $\xi_\mathrm{env}=1.0$, and $\nu_1=\nu_2=1/2$.}
	\label{fig:phase_coexist}
\end{figure}
Figure \ref{fig:phase_coexist} shows both lines for $\varphi=0.01$, $e=0.9$ $\xi_\mathrm{env}=1.0$, and $\nu_1=\nu_2=1/2$, where the critical point is given by $\theta_\mathrm{c}\simeq 34.8$, $\dot\gamma_\mathrm{c}^*\simeq 4.81$, and $(\sigma^{(1)}/\sigma^{(2)})_\mathrm{c}\simeq1.46$.
This means that two (ignited and quenched) states can coexist for $\sigma^{(1)}/\sigma^{(2)}\lesssim 1.46$.
This result is quite analogous to the transition from DST-like to CST-like behaviors for monodisperse cases \cite{Hayakawa17}.

\subsection{Effect of the size ratio for $N_1\neq N_2$}
\begin{figure}[htbp]
	\centering
	\includegraphics[width=0.5\linewidth]{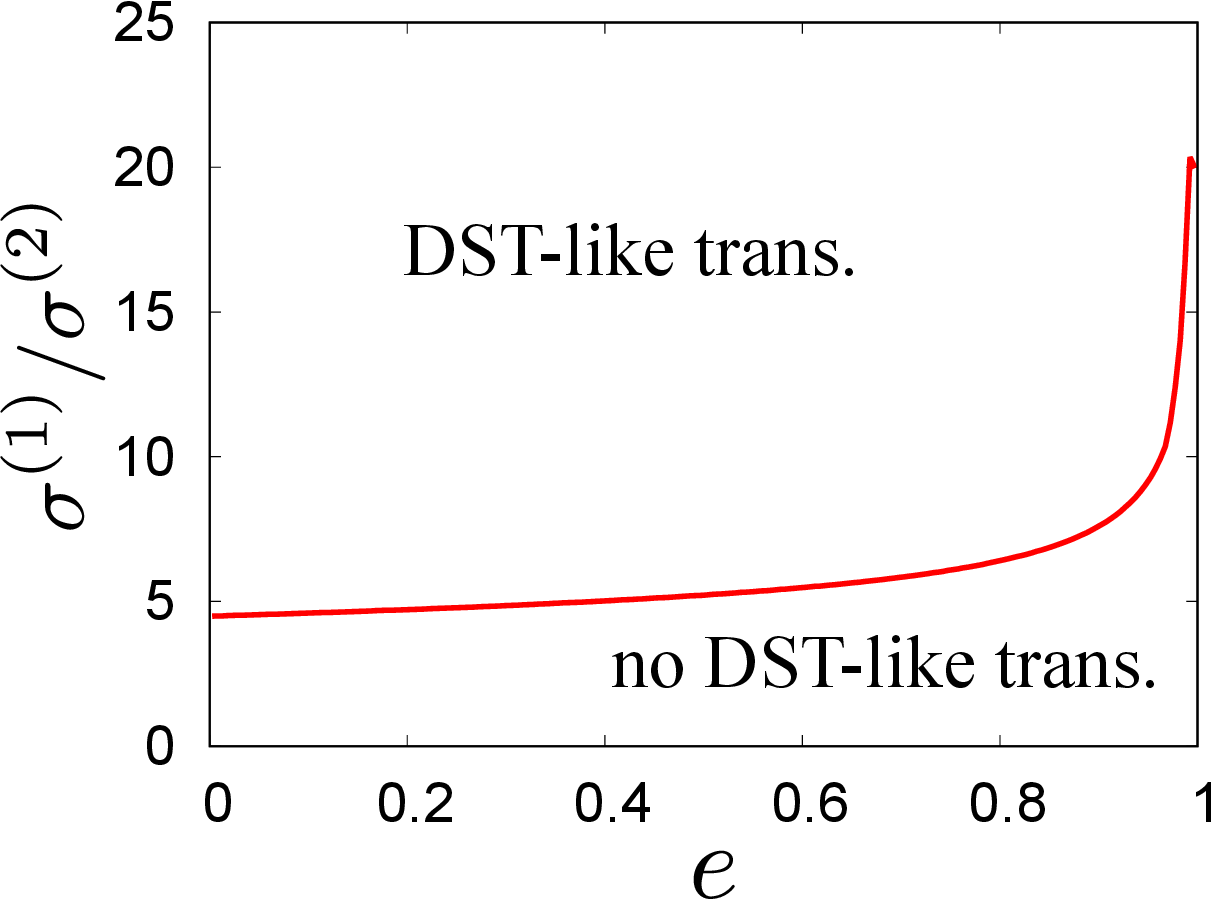}
	\caption{Plot of the critical size ratio against the restitution coefficient $e$ when we fix $\varphi=0.01$, $\xi_{\rm env}=1.0$, and $\mathcal{V}=1$.}
	\label{fig:critical_sigma_ratio}
\end{figure}
Let us consider the case of constant volume ratio $\mathcal{V}=1$.
As shown in Fig.~\ref{fig:N_dependence}, the discontinuous transition occurs as the fraction $\nu_1$ decreases, i.e., the size ratio increases.
This transition is different from the one found in Appendix \ref{sec:size_ratio_N1_N2}.
Figure \ref{fig:critical_sigma_ratio} plots the critical line between the discontinuous transition and continuous transition for $\varphi=0.01$, $\xi_{\rm env}=1.0$, and $\mathcal{V}=1$ based on the theoretical calculation.
As the restitution coefficient $e$ increases, the minimum size ratio also increases, which means that the fraction $\nu_1$ decreases as $\nu_1=1/[1+(\sigma^{(1)}/\sigma^{(2)})^3]$ from the definition of the volume ratio $\mathcal{V}$.
Unfortunately, it is a tough job to check this behavior in simulations.
When the DST-like transition occurs, one needs to simulate a situation where multiple collisions between large particles occur. 
However, as the size ratio increases, the fraction of the larger particles, $\nu_1$, becomes small, then the collision frequency between them also decreases.
This means that the time for multiple collisions exceeds the limit of realistic simulation time. 

\section{Detailed analysis in the tracer limit and the finite size effect of the simulation results}\label{sec:tracer_limit}

In this Appendix, we display the explicit expressions of the partial pressure tensors of a binary mixture in the tracer limit. These expression are then compared with the simulation results when the number of particles is small.

In the tracer limit ($\nu_1\to 0$), the kinetic equation for the velocity distribution function $f_2$ of the excess granular gas 2 is the (closed) nonlinear Boltzmann equation since its state is not perturbed by the presence of the tracer particles 1.
This means that collisions {between tracer and gas particles in the kinetic equation for $P_{\alpha\beta}^{(2)}$ can be neglected, i.e., $\Lambda^{(21)}_{\alpha\beta}+\Lambda^{(22)}_{\alpha\beta}\to\Lambda^{(22)}_{\alpha\beta}$ in Eq.\ \eqref{eq:P_eq} for $i=2$}. {In addition, since the concentration of tracer particles is negligible, one can also neglect the tracer-tracer collisions in the kinetic equation for $P_{\alpha\beta}^{(1)}$. This implies that $\Lambda^{(11)}_{\alpha\beta}+\Lambda^{(12)}_{\alpha\beta}\to\Lambda^{(12)}_{\alpha\beta}$ in Eq.\ \eqref{eq:P_eq} for $i=1$}.

The expressions of the (reduced) elements of the pressure tensor {$\Pi_{\alpha\beta}^{(2)}$} coincide with those obtained for a monodisperse granular suspension. The nontrivial components of {$\Pi_{\alpha\beta}^{(2)}$} are given by \cite{DST16}
\begin{align}
	\Pi_{yy}^{(2)}
	&= -\frac{\lambda_\eta^{(2)*}\sqrt{\theta_2}+2(1-\theta_2^{-1})}{\nu_\eta^{(2)*}\sqrt{\theta_2} + 2},\\
    \Pi_{xy}^{(2)}
	&= -\frac{2\theta_2^{-1}-\left(\lambda_\eta^{(2)*}-\nu_\eta^{(2)*}\right)\sqrt{\theta_2}}{\left(\nu_\eta^{(2)*}\sqrt{\theta_2} + 2\right)^2}
\widetilde{\dot{\gamma}},
\end{align}
where {$\widetilde{\dot{\gamma}}=\dot{\gamma}/\zeta_2$} and we have introduced
\begin{align}
	\lambda_\eta^{(2)*}
	&\equiv \frac{8}{\sqrt{\pi}} (1-e_{22}^2)\varphi_2
	\sqrt{\frac{T_{\rm env}}{m_2\sigma^{(2)2}\zeta_2}},\\
	\nu_\eta^{(2)*}
	&\equiv \frac{24}{5\sqrt{\pi}}(1+e_{22})(3-e_{22}) \varphi_2
	\sqrt{\frac{T_{\rm env}}{m_2\sigma^{(2)2}\zeta_2}},
\end{align}
with the partial volume fraction
\begin{equation}
	\varphi_2
	\equiv \frac{\pi}{6} n \nu_2 \sigma^{(2)3}.
\end{equation}
Here, it should be noted that the global temperature is approximately given by $\theta\simeq \theta_2$ in the tracer limit \cite{Garzo02}.
Using the same procedure {as} in Ref.\ \cite{DST16}, the reduced shear rate $\dot\gamma^*$ is written in terms of the reduced temperature $\theta_2$ as
\begin{equation}
	\dot\gamma^*
	= \left(\nu_\eta^{(2)*}\sqrt{\theta_2}+2\right)
	\sqrt{\frac{3}{2} \frac{\lambda_\eta^{(2)*}\sqrt{\theta_2} + 2(1-\theta_2^{-1})}
		{\left(\nu_\eta^{(2)*}-\lambda_\eta^{(2)*}\right)\sqrt{\theta_2} + 2\theta_2^{-1}}}.
\end{equation}
Now, let us calculate the quantities for the {tracer} species $1$.
First, the quantities $\widetilde{\Lambda}_{\alpha\alpha}^{(12)}$, $\widetilde{\Lambda}_{xy}^{(12)}$, and $\widetilde{\Lambda}_{xy}^{\prime (12)}$ are written as
\begin{subequations}
\begin{align}
    \widetilde{\Lambda}_{\alpha\alpha}^{(12)}
	&\equiv {\frac{1}{m_2^{*3/2}}(\vartheta^\prime+1)^{1/2}
	\left\{\left[m_1^{*}+\frac{1}{2}m_{2}^*\left(1-e_{12}\right)\right]\vartheta^\prime - \frac{1}{2}m_{2}^*(1+e_{12})\right\}},\\
	\widetilde{\Lambda}_{xy}^{(12)}
	&\equiv \frac{1}{10m_1^* m_2^{*1/2}}\frac{1}{\theta_2 (\vartheta^\prime+1)^{1/2}}
		\left[2(5+6\vartheta^\prime) -3\mu_{21}(\vartheta^\prime+1)(1+e_{12})\right],\\
	\widetilde{\Lambda}_{xy}^{\prime (12)}
	&\equiv \frac{1}{10m_2^{*3/2}}\frac{{\theta_2^{-1}}}{(\vartheta^\prime+1)^{1/2}}
		\left[3\mu_{21}(\vartheta^\prime+1)(1+e_{12})-2\vartheta^\prime\right],
\end{align}
\end{subequations}
where we have introduced $\vartheta^\prime\equiv m_2\theta_1/(m_1\theta_2)$.
Then, the nonzero elements of $\Pi_{\alpha\beta}^{(1)}$ {read}
\begin{align}
	\Pi_{xy}^{(1)}&=
	= \frac{3}{\dot{\gamma}^*\theta_1}\left[\zeta_1^*(1-\theta_1)-\frac{1}{2}C_{12}\widetilde{\Lambda}_{\alpha\alpha}^{(12)}\theta_2^{3/2}\right],
\label{eq:Pi_yy1}\\
	\Pi_{xx}^{(1)} &= -2 \Pi_{yy}^{(1)},\\
	\Pi_{yy}^{(1)}&= \Pi_{zz}^{(1)}
	=-\left(1+\frac{2}{\dot{\gamma}^*}\zeta_1^*\Pi_{xy}^{(1)}\right)-2\frac{C_{12}}{\dot{\gamma}^*}\theta_2^{3/2}\left[
    \widetilde{\Lambda}_{xy}^{(12)}\theta_1\Pi_{xy}^{(1)}-\widetilde{\Lambda}_{xy}^{\prime(12)}\theta_2\Pi_{xy}^{(2)}\right].
    \label{eq:Pi_xy1}
\end{align}
Substituting Eqs.\ \eqref{eq:Pi_yy1}--\eqref{eq:Pi_xy1} into Eq.\ \eqref{eq:steady_eq1} with $C_{11}=0$, we can obtain the equation which determines $\theta_1$ as
\begin{equation}
	\frac{2}{3}{\dot{\gamma}^*} \theta_1 \Pi_{xy}^{(1)}
	= 2\zeta_1^*(1-\theta_1) - C_{12}\widetilde{\Lambda}_{\alpha\alpha}^{(12)} \theta_2^{3/2}.
\end{equation}

Figure \ref{fig:tracer_limit} presents the shear-rate dependence of both the temperature ratio $\vartheta$ and the viscosity ratio $\eta_1/\eta_2$ in the tracer limit.
It should be noted that the flow curves become smooth in the whole range of the shear rate even for a larger size ratio.
\begin{figure}[htbp]
	\centering
	\includegraphics[width=\linewidth]{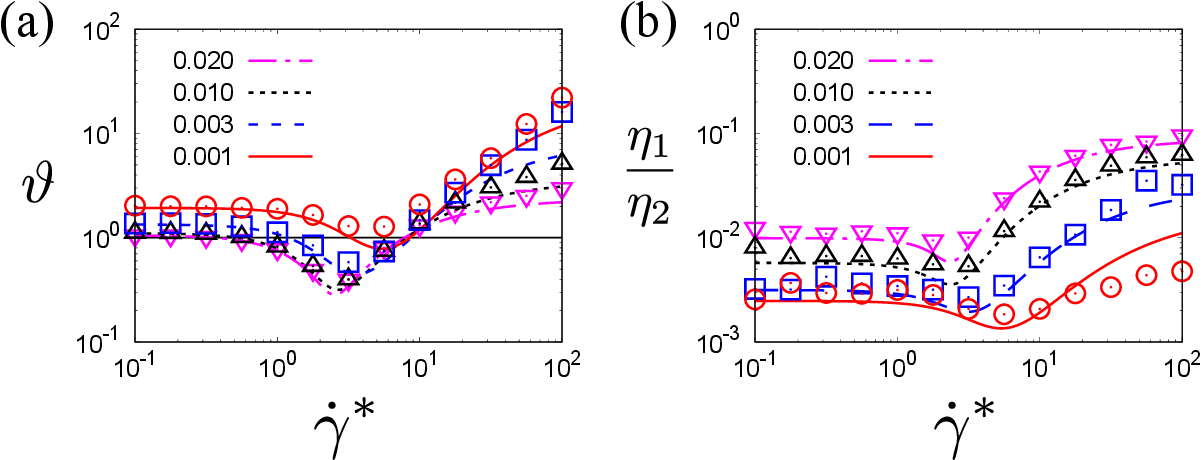}
	\caption{(a) Temperature ratio $\vartheta$ and (b) viscosity ratio $\eta_1/\eta_2$ against the dimensionless shear rate $\dot\gamma^*$ in the tracer limit for the same set of parameters of Fig.\ \ref{fig:vol_ratio0.50}.
	The data of the simulation are obtained for $N=1000$.}
	\label{fig:tracer_limit}
\end{figure}

The limitation of the tracer limit is also understood in Fig.\ \ref{fig:Lambda_ratio}, where the absolute values of the ratio $\left|\Lambda_{\alpha\alpha}^{(ij)}/\Lambda_{\alpha\alpha}^{(22)}\right|$ are plotted as a function of the dimensionless shear rate. Here, the expression of $\Lambda_{\alpha\alpha}^{(ij)}$ is given by Eq.\ \eqref{eq:Lambda_ij_alphabeta}.
In the low shear regime, the values for $(i,j)=(1, 1)$ and $(2, 1)$ are smaller than unity, which means that the contributions {coming from the collisions between the large tracer particles} are negligible.
This indicates that the tracer limit {description is a reasonable approximation} in this regime.
In the high shear regime, on the other hand, the contributions from the collisions between {tracer} particles play an important role to the flow curve, though the number of collisions is small.
Moreover, it is interesting that $\Lambda_{\alpha\alpha}^{(2,1)}$ and $\Lambda_{\alpha\alpha}^{(1,2)}$ become negative in the high and intermediate shear regimes, respectively, though their origins are not clear.
As the number of particles used in the simulation increases, the simulation results recover the {values of $\vartheta$} and $\eta_1/\eta_2$ in the high shear regime.
Then, we expect that the results of simulation for $N \to \infty$ {agree with} the theoretical results.
{In other words}, the results of EDLSHS containing a small number of particles is not reliable in this regime.

\begin{figure}[htbp]
	\centering
	\includegraphics[width=0.5\linewidth]{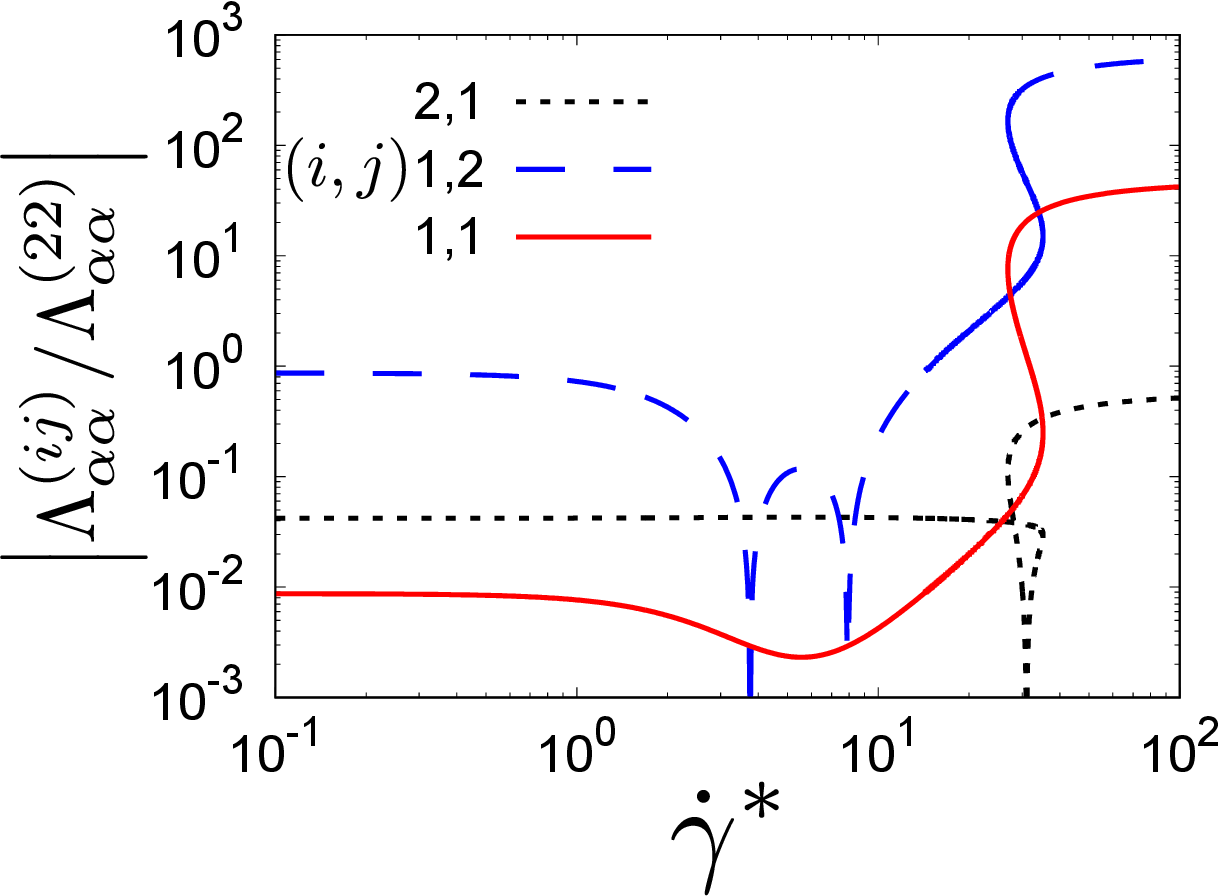}
	\caption{Plot of $\left|\Lambda_{\alpha\alpha}^{(ij)}/\Lambda_{\alpha\alpha}^{(22)}\right|$ for $(i,j)=(1, 1)$ (solid line), $(1, 2)$ (dashed line), and $(2, 2)$ (dotted line) against the dimensionless shear rate $\dot\gamma^*$ for $\varphi=0.01$, $\xi_{\rm env}=1.0$, $e=0.9$, $\mathcal{V}=1$, and $\nu_1=1.0\times10^{-3}$.}
	\label{fig:Lambda_ratio}
\end{figure}

\section{Two-dimensional velocity distribution function of BGK model}\label{sec:BGK}

A possible way of overcoming the mathematical difficulties associated with the Boltzmann collision operators $J_{ij}[f_i,f_j]$ is to use a kinetic model.
As usual, the idea behind a kinetic model is to replace the true operator $J_{ij}$ by a
simpler term that retains the main physical properties of the above operator. In the case of dilute granular mixtures, a BGK-like kinetic model was proposed {in} Ref.~\cite{VegaReyes07}.
In the case of the USF state {(where $\bm{U}_1 = \bm{U}_2=\bm{U}$)}, the BGK-like model is obtained by the replacement of the Boltzmann collision operator $J_{ij}[f_i,f_j]$ by the diffusive term
\begin{equation}
	J_{ij}[\bm{V}|f_i,f_j]
	\to -\frac{1+e_{ij}}{2\tau_{ij}}(f_i-f_{ij})+\frac{\varepsilon_{ij}}{2}\frac{\partial}{\partial\bm{V}}\cdot (\bm{V}f_i),
	\label{G_2.1}
\end{equation}
where we have introduced the quantities
\begin{align}\label{G_2.2}
	\frac{1}{\tau_{ij}}
	&=\frac{8\sqrt{\pi}}{3}n_j\sigma^{(ij)2} {\left(\frac{2T_i}{m_i}+\frac{2T_j}{m_j}\right)^{1/2}},\\
	\varepsilon_{ij}&=\frac{1}{2\tau_{ij}}\frac{m_{ij}^2}{m_i^2}\left(1+\frac{m_iT_j}{m_jT_i}\right)(1-e_{ij}^2), \\
	f_{ij}(\bm{V})&=n_i\left(\frac{m_i}{2\pi T_{ij}}\right)^{3/2}\exp\left(-\frac{m_i V^2}{2T_{ij}} \right),\\
	T_{ij}&=T_i+\frac{2m_im_j}{(m_i+m_j)^2}(T_j-T_i).
\end{align}

The corresponding BGK-like equation for the distribution $f_1$ in the steady USF is
\begin{equation}
	-\dot\gamma V_y\frac{\partial f_1}{\partial V_x}
	- \zeta_1 \frac{\partial}{\partial\bm{V}}\cdot (\bm{V}f_1)
	-\frac{\zeta_1 T_{\rm env}}{m_1} \frac{\partial^2f_1}{\partial V^2}
	=-\frac{1}{2}\sum_{j=1}^2
	\left[
		\frac{1+e_{1j}}{\tau_{1j}}(f_1-f_{1j})-\varepsilon_{1j} \frac{\partial}{\partial\bm{V}}\cdot (\bm{V}f_1)
	\right].
	\label{G_2.4}
\end{equation}
The kinetic equation for $f_2$ is obtained from Eq.\ \eqref{G_2.4} by setting $1\leftrightarrow 2$.
So far, we have not {been able to obtain} the explicit exact form of $f_i(\bm{V})$ in Eq.\ \eqref{G_2.4}.
An exception corresponds to the simple limit case $T_{\rm env} = 0$ with keeping $\zeta_i = {\rm const}$.
It corresponds to a situation {where} the background temperature $T_{\rm env}$ is much smaller than the kinetic temperature $T$ under the high shear rate limit.
Hence, the suspension model ignores the effects of thermal fluctuations on solid particles and the impact of the gas phase on grains is only accounted by the drag force term.
Although $\zeta_i$ should be proportional to $\sqrt{T_{\rm env}}$ for hard-core molecules, such a simplified model has been employed in some previous works~\cite{Tsao95,Sangani96,Saha17}.

If we take the limit $T_{\rm env}/T\ll 1$, Eq.\ \eqref{G_2.4} becomes
\begin{equation}
	-\dot\gamma V_y \frac{\partial f_1}{\partial V_x}-3\alpha_1 f_1
	-\alpha_1\bm{V}\cdot \frac{\partial f_1}{\partial \bm{V}} + \xi_1 f_1
	= \Phi_1,\label{G_2.5}
\end{equation}
where we have introduced the parameters:
\begin{align}
	\alpha_1&=\zeta_1+\frac{\varepsilon_{11}+\varepsilon_{22}}{2},\\
	\xi_1&=\frac{1}{2}\Bigg(\frac{1 + e_{11}}{\tau_{11}} + \frac{1 + e_{12}}{\tau_{12}}\Bigg), \\
	\Phi_1&=\frac{1}{2}\Bigg(\frac{1 + e_{11}}{\tau_{11}}f_{11}+\frac{1 + e_{12}}{\tau_{12}}f_{12}\Bigg).
\end{align}
The formal solution of Eq.~\eqref{G_2.5} can be written as
\begin{align}
	f_1(\bm{V})
	&=\left(\xi_1-3\alpha_1-\dot\gamma V_y \frac{\partial}{\partial V_x}
		-\alpha_1 \bm{V}_1\cdot \frac{\partial}{\partial \bm{V}_1}\right)^{-1}\Phi_1(\bm{V}) \nonumber\\
	&=\int_0^\infty ds \; e^{-(\xi_1-3\alpha_1)s}
	e^{\dot\gamma s V_y \frac{\partial}{\partial V_x}}
	e^{\alpha_1 s \bm{V}\cdot \frac{\partial}{\partial \bm{V}}} \Phi_1(\bm{V}). \label{G_2.9}
\end{align}
Note that the velocity operators {appearing} in Eq.~\eqref{G_2.9} commute.
Their action on an arbitrary function $g(\bm{V})\equiv g(V_x,V_y,V_z)$ is
\begin{subequations}
\begin{align}
	e^{\dot\gamma s V_y \frac{\partial}{\partial V_x}} g(\bm{V})
	&=g(V_x+\dot\gamma s V_y,V_y,V_z), \label{G_2.10}\\
	e^{\alpha_1 s \bm{V}\cdot \frac{\partial}{\partial \bm{V}}}g(\bm{V})
	&=g\left(e^{\alpha_1s}\bm{V}\right). \label{G_2.11}
\end{align}
\end{subequations}

Taking into account the action of these operators in Eq.\ \eqref{G_2.9}, the velocity distribution $f_1(\bm{V})$ can be written as
\begin{equation}
	f_1(\bm{V}) = n_1 \left(\frac{m_1}{2T_1}\right)^{3/2} g_{1, {\rm B}}(\bm{c}),\quad
	\bm{c}\equiv \left(\frac{m_1}{2T_1}\right)^{-1/2} \bm{V},
\end{equation}
where
\begin{align}
	g_{1, {\rm B}}(\bm{c})
	&= \pi^{-3/2} \int_0^\infty ds\; e^{-(\xi_1-3\alpha_1)s}
	\left\{\frac{1+e_{11}}{2\tau_{11}} \chi_1^{-3/2}
		\exp\left[-\chi_1^{-1} e^{2\alpha_1 s}\left((c_x+\dot\gamma s c_y)^2+c_y^2 + c_z^2\right)\right]\right.\nonumber\\
	&\left.\hspace{12.5em}+\frac{1+e_{12}}{2\tau_{12}}\chi_{12}^{-3/2}
		\exp\left[-\chi_{12}^{-1} e^{2\alpha_1 s}\left((c_x+\dot\gamma s c_y)^2+c_y^2 + c_z^2\right)\right]\right\}\nonumber\\
	&= \pi^{-3/2} \int_0^\infty ds \; e^{-(\xi_1^*-3\alpha_1^*)s}
	\left\{\frac{1+e_{11}}{2\tau_{11}^*} \chi_1^{-3/2}
		\exp\left[-\chi_1^{-1} e^{2\alpha_1^* s}\left((c_x+\dot\gamma^* s c_y)^2+c_y^2 + c_z^2\right)\right]\right.\nonumber\\
	&\hspace{12.5em}\left.+\frac{1+e_{12}}{2\tau_{12}^*}\chi_{12}^{-3/2}
		\exp\left[-\chi_{12}^{-1} e^{2\alpha_1^* s}\left((c_x+\dot\gamma^* s c_y)^2+c_y^2 + c_z^2\right)\right]\right\}. \label{G_2.13}
\end{align}
Here, $\xi_1^*\equiv\xi_1\overline{\sigma}/\sqrt{\overline{m}/T_{\rm env}}$, $\alpha_1^*\equiv \alpha_1\overline{\sigma}/\sqrt{\overline{m}/T_{\rm env}}$, $\tau_{ij}^*\equiv \tau_{ij}\sqrt{T_{\rm env}/\overline{m}}/\overline{\sigma}$, $\chi_1\equiv T_1/T$, and $\chi_{12}\equiv T_{12}/T$.

To illustrate the shear-rate dependence of the BGK distribution $g_{1, {\rm B}}(\bm{c})$, let us define the marginal (two-dimensional) distribution function
\begin{equation}
	g_{1, {\rm B}}^{(xy)}(c_x, c_y) = \int_{-\infty}^\infty dc_z g_{1, {\rm B}}(\bm{c}).
\end{equation}
From Eq.\ \eqref{G_2.13}, one gets
\begin{align}
	g_{1, {\rm B}}^{(xy)}(c_x, c_y)
	&= \frac{1}{\pi} \int_0^\infty ds \; e^{-(\xi_1^*-2\alpha_1^*)s}
	\left\{\frac{1+e_{11}}{2\tau_{11}^*} \chi_1^{-1} \exp\left[-\chi_1^{-1}e^{2\alpha_1^* s}
	\left((c_x+\dot\gamma s c_y)^2+c_y^2\right)\right]\right.\nonumber\\
	&\hspace{11em}\left. + \frac{1+e_{12}}{2\tau_{12}^*} \chi_{12}^{-1}
	\exp\left[-\chi_{12}^{-1}e^{2\alpha_1^* s}\left((c_x+\dot\gamma s c_y)^2+c_y^2\right)\right]\right\}.
	\label{eq:g_1xy_BGK}
\end{align}

Figure \ref{fig:prob_2.0_e0.9_BGK} shows how this model works when we compare with the simulation results.
Interestingly, the BGK-like model gives the correct VDF in the wider range of $(c_x, c_y)$ plane in the high shear regime.
In particular, {some features of the true VDF (such as the enhancement in the shear direction and the form of $g_1^{(xy)}$ near the positive and negative peaks) are captured in a more precise way by the BGK distribution than the Grad's distribution} (see Figs.\ \ref{fig:prob_2.0_e0.9} and \ref{fig:prob_2.0_e0.9_BGK}). Nevertheless, {we recall that the applicability of the solution Eq.\ \eqref{eq:g_1xy_BGK} to the BGK-like model is limited to the high shear regime}.
As the environmental temperature plays a role in the rheology, {the BGK solution \eqref{eq:g_1xy_BGK}} cannot capture the properties of the VDF in the complete range of shear rates (see Figs.\ \ref{fig:prob_2.0_e0.9_BGK}(a), (b), and (c)).

\section{One-dimensional velocity distribution function}\label{sec:1D_VDF}

In the main text and Appendix \ref{sec:BGK}, we have compared the marginal two-dimensional velocity distribution function obtained from the simulations {with those} obtained from Grad's method and the BGK-like model. In this Appendix, on the other hand, we investigate whether both approximations work when we consider the one-dimensional velocity distribution.

Let us define the marginal (one-dimensional) distribution obtained from Grad's method
\begin{equation}
	g_{1,{\rm G}}^{(x)}(c_x)
	\equiv \int_{-\infty}^\infty dc_y g_{1, {\rm G}}^{(xy)}(c_x, c_y)
	=\frac{1}{\sqrt{\pi}} e^{-c_x^2}\left(1-\frac{\Pi_{xx}^{(1)}}{2}+\Pi_{xx}^{(1)}c_x^2\right),
	\label{eq:g1_Grad_1D}
\end{equation}
and {from the BGK model}
\begin{align}
	&g_{1,{\rm B}}^{(x)}(c_x)
	\equiv \int_{-\infty}^\infty dc_y g_{1, {\rm B}}^{(xy)}(c_x, c_y)\nonumber\\
	&= \frac{1}{\sqrt{\pi}} \int_0^\infty ds
	\frac{e^{-(\xi_1^*-\alpha_1^*)s}}{\sqrt{1+\dot\gamma^{*2}s^2}}
	\left[\frac{1+e_{11}}{2\tau_{11}^*} \chi_1^{-1/2} \exp\left(-\chi_1^{-1}
		e^{2\alpha_1^* s}\frac{c_x^2}{1+\dot\gamma^{*2}s^2}\right)\right.\nonumber\\
	&\hspace{12em}\left.
	+ \frac{1+e_{12}}{2\tau_{12}^*} \chi_{12}^{-1/2} \exp\left(-\chi_{12}^{-1}
		e^{2\alpha_1^* s}\frac{c_x^2}{1+\dot\gamma^{*2}s^2}\right)\right].
	\label{eq:g1_BGK_1D}
\end{align}

Figure \ref{fig:1D_VDF} shows the comparison of the VDF obtained from the simulations with Eqs.\ \eqref{eq:g1_Grad_1D} and \eqref{eq:g1_BGK_1D} when we control the shear rate from $\dot\gamma^*=0.32$ to $32$.
Here, we have fixed $\varphi=0.01$, $e=0.9$, $\xi_{\rm env}=1.0$, and $\sigma^{(1)}/\sigma^{(2)}=2.0$.
The one-dimensional VDF estimated from Grad's method works well in the wide range of the shear rate, although this approximation cannot reproduce the fat tail of the VDF in the intermediate regime.
The consistency in the high shear regime is different when we compare with the two-dimensional VDF in Figs.\ \ref{fig:prob_2.0_e0.9}(c), (d), and (e).
On the other hand, the BGK one-dimensional VDF is worse than that of Grad's distribution.
Although it captures the behavior of the VDF near $c_x\sim 0$, the {solution to the} BGK-like model overestimates the high energy tail of the VDF in the high shear regime (see Fig.\ \ref{fig:1D_VDF}(d) and (e)).

\begin{figure}[htbp]
	\centering
	\includegraphics[width=\linewidth]{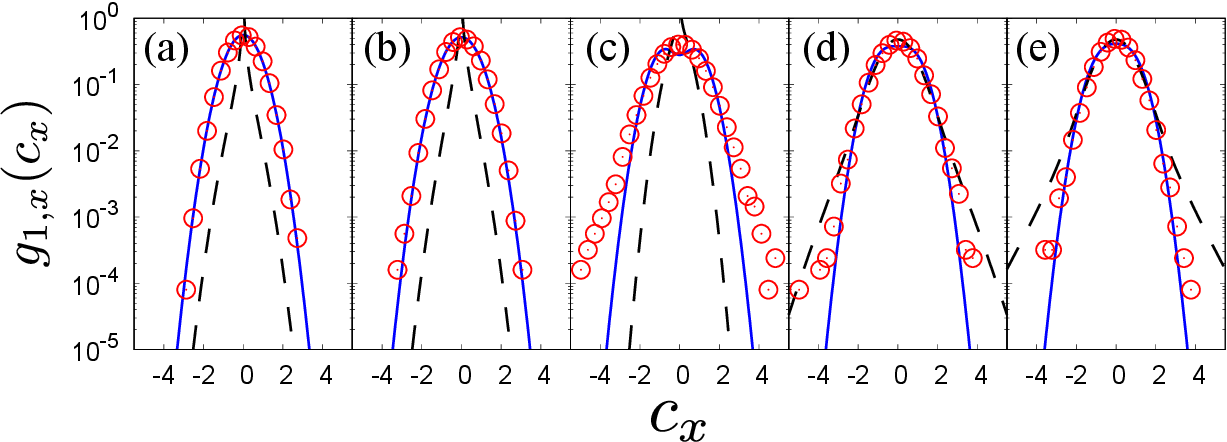}
	\caption{One-dimensional velocity distribution functions of the larger particles for (a) $\dot\gamma^*=0.32$, (b) $1.0$, (c) $3.2$, (d) $10$, and (e) $32$ when we fix $\varphi=0.01$, $e=0.9$, $\xi_{\rm env}=1.0$, and $\sigma^{(1)}/\sigma^{(2)}=2.0$.
	The solid and dashed lines represent Grad's approximation \eqref{eq:g1_Grad_1D} and the BGK model \eqref{eq:g1_BGK_1D}, respectively.}
	\label{fig:1D_VDF}
\end{figure}

\let\doi\relax

\end{document}